# Lithium Intercalation in the Anisotropic van der Waals Magnetic Semiconductor CrSBr

*Kseniia Mosina[1,†], Aljoscha Söll[1,†], Jiri Sturala[1,\*], Martin Veselý[2,\*], Petr Levinský[3], Florian Dirnberger[4], Giuliana Materzanini[5], Nicola Marzari[6], Gian-Marco Rignanese[5,7], Borna Radatović[1], David Sedmidubsky[1], and Zdeněk Sofer[1,\*]*

[1] Department of Inorganic Chemistry, University of Chemistry and Technology Prague, Technická 5, 166 28, Prague 6, Czech Republic

[2] Department of Organic Technology, University of Chemistry and Technology Prague, Technická 5, 166 28, Prague 6, Czech Republic

[3] Institute of Physics of the Czech Academy of Sciences, Cukrovarnická 10, 162 00, Prague 6, Czech Republic

[4] Department of Physics, TUM School of Natural Sciences, Technical University of Munich, Garching, Germany

[5] Institute of Condensed Matter and Nanosciences (IMCN), Université catholique de Louvain, Louvain-la-Neuve 1348, Belgium

[6] Theory and Simulation of Materials (THEOS), and National Centre for Computational Design and Discovery of Novel Materials (MARVEL), École Polytechnique Fédérale de Lausanne (EPFL), CH-1015 Lausanne, Switzerland

[7] Wel Research Institute, Wavre, Belgium

† These authors contributed equally.

## Abstract

Alkali metal intercalation is an important strategy for doping van der Waals materials. Lithium, in particular, was shown to achieve exceptional charge carrier densities, reaching levels at which fundamental electrical, optical, and magnetic material properties begin to be strongly modified. While lithium is known to be highly volatile, its migration dynamics in anisotropic layered crystals remain poorly understood. In this work, we investigate the intercalation of lithium in-between layers of the anisotropic magnetic semiconductor CrSBr. Using exfoliated crystals, we are able to monitor the dynamics of the intercalation process in real time through optical and electrical characterization methods. Our measurements reveal highly anisotropic migration of Lithium characterized by diffusion coefficients that differ by more than one order of magnitude along *a*- and *b*-directions. This finding is in good agreement with our molecular dynamics simulations which show trajectories of lithium atoms primarily follow the Br-chains in the *a*-direction. Beyond that, we find that partially covering CrSBr crystals by thin hexagonal boron nitride (hBN) flakes has a significant impact on the intercalation process, and that lithium strongly enhances the electrical conductivity along the *a*-axis. Our method offers a new platform for lithium diffusion studies and encourages further research to pursue the fabrication of lithium-doped devices.

# Introduction

Lithium intercalation is one of the most powerful approaches for vastly tuning charge doping in layered semiconductors, magnets, and superconductors, and is traditionally accomplished via electrochemical methods.[1–4] Electrochemical intercalation plays a central role in many liquid phase exfoliation protocols since it often induces bulk crystal delamination down to exfoliated flakes of high crystalline quality.[5–8] However, the fact that it relies on complex cell architectures limits in situ studies of intercalation into individual flakes. In contrast, chemical intercalation exploits the chemical potential of the intercalant as a driving force, bypassing the need for a working electrode or cell. This enables direct observation of individual flakes lithiation on standard substrates (e.g., $SiO_2$/Si) and facilitates in situ optical and spectroscopic characterization.[9] Since the 1960s, researchers have used chemical intercalation to explore how the impact of lithium intercalation on the electrical and magnetic properties of layered semiconductors such as FeOCl.[10–12]

Another promising candidate for the study of lithium intercalation is the van der Waals magnetic semiconductor CrSBr. In its bulk state CrSBr exhibits a direct band gap of 1.5 eV and an orthorhombic crystal structure of the *Pmmn* space group and $D_{2h}$ point group.[13] Each layer of CrSBr consists of covalently bonded network of Cr and S atoms, while Br atoms are bound exclusively to Cr atoms and arranged in chains along the a-direction (Figure 1a).[14] This distinctive atomic arrangement leads to significant structural anisotropy, evidenced by distinct lattice constants along the crystallographic axes (a = 3.506 Å, b = 4.767 Å, and c = 7.965 Å).[14] The relatively weak interlayer hybridization facilitates an easy cleavage of layers when exfoliating the material, resulting in rectangular sheets elongated in the *a*-direction due to the in-plane bonding anisotropy. CrSBr has garnered considerable attention in the past few years because of its anisotropic electronic transport and distinctive magnetic behavior with A-type antiferromagnetic ordering below $T_N \approx 132$ K and ferromagnetism in the monolayer.[15,16] Feuer et al. recently reported that Li-CrSBr can be obtained through chemical intercalation with lithium naphthalenide, and it exhibits ferromagnetic ordering with a $T_C \approx 200$ K.[17] They further observe the emergence of a charge density wave state and demonstrate a substantial increase in conductivity in the *a*-direction due to charge carrier doping. While all studies in the aforementioned work were conducted on intercalated bulk material, the importance of studying exfoliated samples was also highlighted. For example, thin flakes would allow direct transport measurements in the *b*-direction that are hindered by grain boundary formation in bulk Li-CrSBr.[17]

Here, we identify optimal conditions for lithium ions intercalation by monitoring its spatiotemporal dynamics in exfoliated CrSBr crystals in situ using Raman spectroscopy and electrical conductivity measurements. Additionally, we show that partial or complete hBN encapsulation strongly affects the dynamics and in-plane anisotropy of $Li^+$ diffusion. Conductivity measurements further confirm the feasibility of studying charge transport both in the *a* and *b*-directions, thereby paving the way for future device applications leveraging intercalation induced tunability.

## Discussion

## Lithium Intercalation in Exfoliated CrSBr

We present a lithium intercalation method for exfoliated CrSBr, schematically illustrated in Figure 1b-c. To begin, we prepare the sample of mechanically exfoliated CrSBr transferred onto an Si/SiO$_2$ wafer using the conventional Scotch tape method (see Methods) and characterize CrSBr flakes by optical microscopy, atomic force microscopy (AFM), and Raman spectroscopy prior to intercalation. Optical contrast of the flakes is correlated with AFM-measured thickness, and the characteristic $A_g^1, A_g^2, A_g^3$ vibrational modes of CrSBr at 114 cm$^{-1}$, 244 cm$^{-1}$, 342 cm$^{-1}$, respectively, are confirmed via Raman spectroscopy (see Supplementary Information Section 1). To initiate intercalation, the sample is immersed in a lithium anthracenide solution in tetrahydrofuran (THF), as shown in Figure 1c. Further details on solution preparation are provided in the Methods and Figure S2. During the intercalation process, anthracene in its reduced ionic form donates one electron to the CrSBr structural unit, reducing chromium from the formal oxidation state Cr(III) to Cr(II). As the structure must remain charge-neutral, the excess electrons are compensated by positively charged lithium cations intercalating into CrSBr. After intercalation, the sample of Li-CrSBr is rinsed with dry THF and characterized by Raman spectroscopy again, which serves as an effective tool for real-time monitoring of lithiation. To confirm that lithium is indeed intercalated between van der Waals layers rather than accumulating on the surface, we employ time-of-flight secondary-ion mass spectrometry (TOF-SIMS). The technique described in more detail in Supplementary Information Section 2 demonstrates that after intercalation, lithium ions are distributed across CrSBr layers.

To identify suitable intercalation conditions, we test a range of lithium anthracenide concentrations (see Supplementary Information Section 3). We find that a minimum concentration of 0.2 mM is required to initiate intercalation within 15 minutes. Increasing the concentration further accelerates the process but risks damaging thin flakes. Signs of deterioration can be observed at 0.64 mM (Figure S8). A concentration of 0.32 mM thus offers an optimal balance between intercalation efficiency and preservation of the crystal structure. In the experiment, this corresponds to a concentration at which the initially blue lithium anthracenide solution turns colorless during the dilution with dry THF. Hence, it is straightforward to prepare.

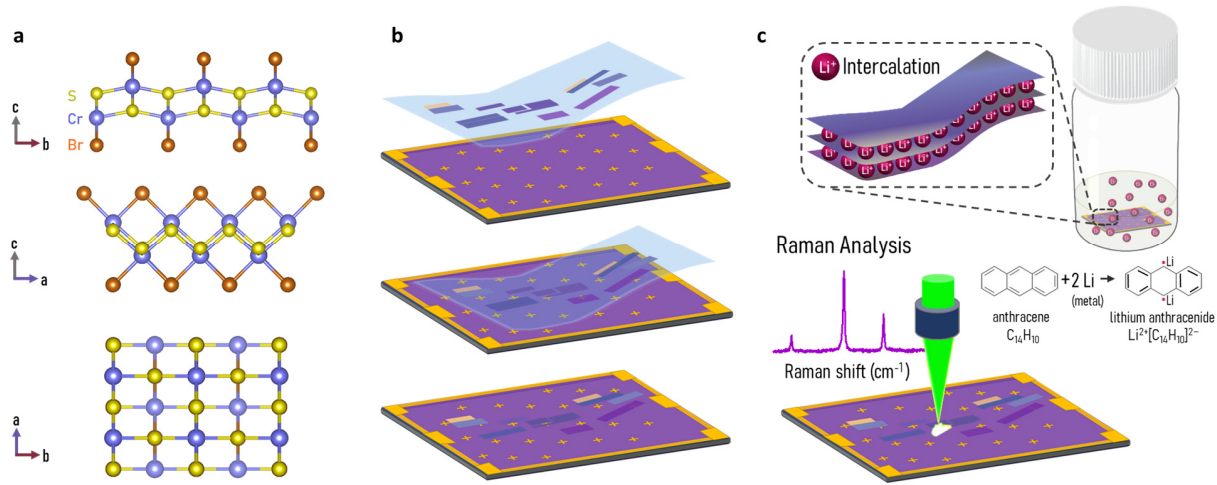

**Figure 1.** Atomic model of CrSBr (a) with projection along the *a*-direction, *b*-direction and *c*-direction. Structures are generated by VESTA software.[18] Schematic illustration of sample preparation starting with the mechanical exfoliation of CrSBr (b). Image (c) shows the process of lithium intercalation into exfoliated CrSBr together with the schematic representation of lithium anthracenide formation in the presence of THF. The wafer with intercalated flakes is retrieved for Raman analysis.

To investigate the effect of intercalation on vibrational Raman modes, we repeatedly intercalate a ~6 nm CrSBr flake with lithium anthracene solution (0.32 mM in THF) in three-minute intervals, acquiring Raman spectra and optical images after each interval, as presented in Figure 2a-b. The resulting spectra reflect changes in the lattice parameters, driven by the insertion of lithium cations into CrSBr. Upon intercalation, Raman spectra reveal two distinct sets of phonon modes, $A_g$ and $A_1$, which is indicative of the presence of both intercalated and non-intercalated material within the area we probe. As intercalation proceeds, $A_g$ modes gradually diminish and new $A_1$ modes emerge at lower frequencies, reflecting the reduction in symmetry from $D_{2h}$ to $C_{2v}$ in Li–CrSBr, as shown in Figure S9.

Within the first three minutes, the onset of $A_1^1$ and $A_1^2$ mode respectively appear at 100 cm$^{-1}$ and at about 220–230 cm$^{-1}$. The $A_g^3$ mode broadens slightly with an asymmetric low-energy tail. At 6 minutes, the $A_1^1$ mode intensity increases, while the $A_g^1$ mode of pristine CrSBr softens yet persists at 114 cm$^{-1}$, and further broadening is observed of both $A_g^2$ and $A_g^3$ peaks. After nine minutes, the $A_1^1$ mode becomes dominant, and a broadened $A_1^2$ feature reflects evolving phonon interactions associated with Li-CrSBr. At 12 minutes, intercalation is complete, as all $A_g$ modes from the pristine material vanish, and $A_1$ modes are fully developed. Extending the time to 15 minutes yields no further changes, indicating completion of lithiation in the 6 nm CrSBr flake within 12 minutes. Despite these spectral changes, the optical contrast of the flake remains unchanged (Figure 2a). We repeat the experiment with another flake to show how the intensity of each mode depends on the laser polarization (Figure 2c). The pristine flake shows the expected polarization dependence for CrSBr with the $A_g^1$ and $A_g^3$ mode most intense for excitation polarization parallel to the *b*-direction E∥*b* and the $A_g^2$ mode most intense for E∥*a*. After intercalation, this trend is mostly maintained, however, $A_1^2$ and $A_1^3$ show a weaker dependence and their intensity does not decrease all the way to zero, which could be attributed to a disorder starting to appear within the structure.

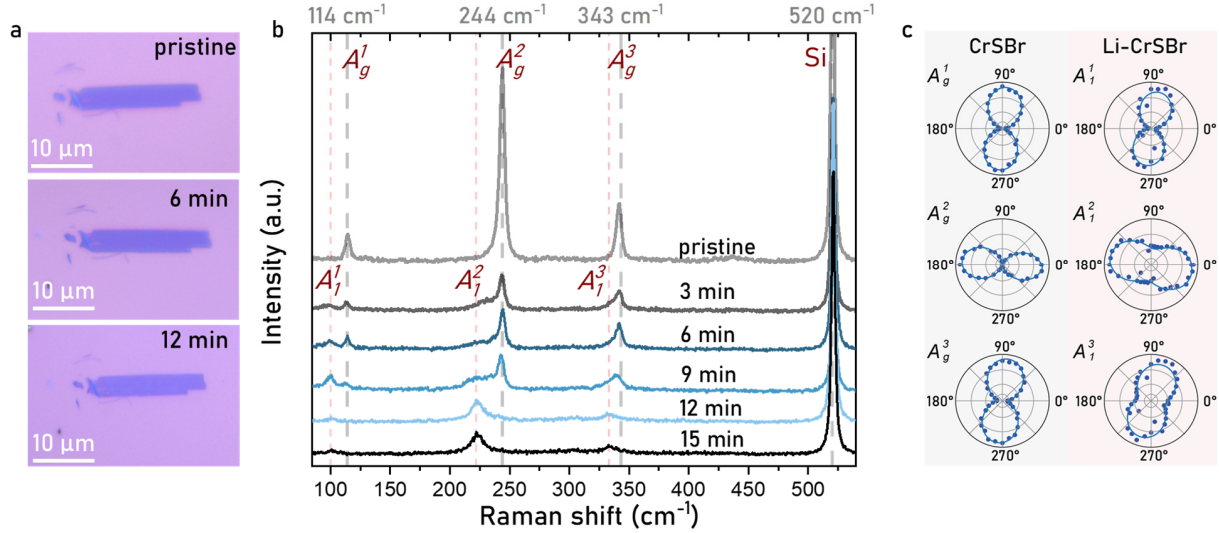

**Figure 2.** (a) Optical micrographs of exfoliated flakes upon intercalation evolution. (b) Corresponding time-resolved Raman spectra of CrSBr flake collected upon intercalation. (c) Polarization dependence of Raman modes for a pristine and an intercalated flake. The dependence becomes less pronounced while the general trend is maintained.

To support our findings in the Raman study, we perform density functional theory (DFT) calculations (see Methods for further details). In pristine CrSBr, the calculated $A_g^1, A_g^2$, and $A_g^3$ modes align well with experimental values. Upon intercalation, symmetry and peak assignment change, resulting in structure CrSBrLi$_{0.5}$ similar to FeOClLi$_x$.[10] The theoretical positions of the $A_1^1, A_1^2, A_1^3$ modes show good agreement with experiment (see Table 1). Atomic displacements are shown in Figure S9. The $A_1^3$ mode is the least affected by lithium intercalation, since that peak originates from Cr and S displacements and its position is not determined with high precision computationally.

Table 1. Calculated and observed Raman shifts related to displacement of selected atoms.

|  | CrSBr | | | Li-CrSBr | |
| --- | --- | --- | --- | --- | --- |
|  | Theory / cm$^{-1}$ | Exp. / cm$^{-1}$ |  | Theory / cm$^{-1}$ | Exp. / cm$^{-1}$ |
| $A_g^1$ | 116.2 | 114 | $A_1^1$ | 92.3 | 95 |
| $A_g^2$ | 241.0 | 244 | $A_1^2$ | 206.5 | 219 |
| $A_g^3$ | 332.1 | 342 | $A_1^3$ | 328.6 | 331 |

To get a better understanding of the intercalation process, we examine the intercalation of a CrSBr flake with a thickness of about 50 nm. The sample is intercalated in the same manner as described above and the progression of intercalation over time is monitored. We track the spatial distribution of the $A_1^1$ mode signal intensity at 95 cm$^{-1}$ and compare it directly with the $A_g^1$ mode at 114 cm$^{-1}$.

Figure 3 shows the optical images during lithiation, along with corresponding peak intensity maps of $A_g^1$ and $A_1^1$ modes. The optical image of the pristine flake is shown in Figure 3a. As expected, the $A_1^1$ mode of Li–CrSBr is absent, while the $A_g^1$ mode of CrSBr shows a strong, uniform signal (Figure 3d). After nine minutes, the $A_1^1$ mode emerges prominently, visualized by high-intensity stripes along the flake edges, while the $A_g^1$ signal weakens, especially at the boundaries, indicating edge-initiated, partial intercalation, which slowly spreads inwards (Figure 3e). After 60 minutes, the intercalated region is clearly visible in the optical micrograph (dark blue stripes in Figure 3c), and the $A_1^1$ mode dominates the Raman response (Figure 3f), confirming progressive lithiation of the flake.

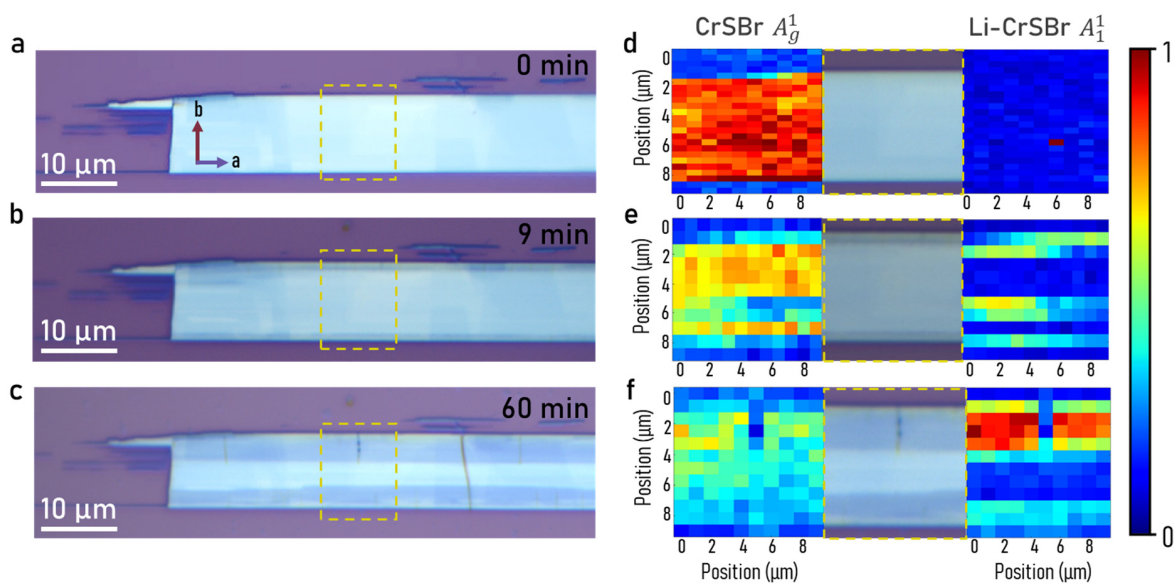

**Figure 3.** Optical micrographs captured at (a) 0 min, (b) 9 min, and (c) 60 min of intercalation. The yellow rectangle in each optical image indicates the specific area chosen for mapping. (d-f) The peak intensity maps of $A_g^1$ and $A_1^1$ modes collected at corresponding time intervals.

At first glance, our observations presented in Figure 3 seem to suggest that intercalation occurs exclusively along the *b*-axis. From a microscopic point of view, however, one could argue that the anisotropic atomic arrangement of Cr-Br chains along the *a*-direction creates a unique bonding situation[13], which should facilitate an easy diffusion of guest ions along the *a*-direction. It lowers the migration energy barrier, creating diffusion channels within the structure, which should dictate the diffusion mechanism. Therefore, we propose that lithium ions are unable to enter the material from the short edge (perpendicular to the *a*-direction). Instead, they enter from the long edge (perpendicular to the *b*-direction), where they rapidly spread through the diffusion channels oriented along the *a*-direction, and the intercalation only gradually progresses inward along the *b*-direction, leading to the observations in Figure 3a-c.

We test this hypothesis in a series of intercalation experiments with fully or partially hBN encapsulated CrSBr. Initially, we confirm that fully encapsulating pristine CrSBr in hBN protects it from intercalation entirely (Figure S10a-c). Then, we cover the CrSBr partly either along the *a* or *b*-direction, leaving one part exposed for lithium entry. When hBN coverage leaves an exposed edge perpendicular to the *b*-axis, intercalation progresses slowly inward along the *b*-direction from this exposed edge, similar to a non-encapsulated flake (Figure S11b). In contrast, when the exposed edge is perpendicular to the *a*-axis, lithium enters and then quickly spreads underneath the hBN covered area along the *a*-direction (Figure S10d-f), suggesting that Li$^+$ diffusion is indeed significantly faster along the *a*-direction.

To get an estimate for the anisotropy ratio of diffusion coefficients, $D_a/D_b$, we partially encapsulate a sheet of CrSBr with a thickness of about 25 nm in hBN along the *a*-direction, as shown in Figure 4b, and intercalate it in multiple 30 s to 60 s intervals. After each intercalation step, we track the progress of intercalation deep within the hBN covered area based on the optical contrast. Figure 4a shows how diffusion progresses for a selected area. Under the assumption of 1D diffusion in each direction, we can gain an estimate for the diffusion coefficient according to $L^2 = 2Dt$.[19–21] In Figure 4c, we present the progression of intercalation in the selected area as $L^2$ vs. $t$ for one diffusion length measurements in each direction for each time step (further details are found in Supplementary Information Section 5). We notice immediately that intercalation along the *a*-direction proceeds markedly faster, as evidenced by the steeper slope. Additionally, we observe two distinct regimes, a slow initial diffusion followed by accelerated diffusion, which is likely triggered by a lattice parameter change that facilitates faster ion transport. We calculate an average anisotropy ratio of diffusion coefficients of $D_a/D_b \approx 21.8 \pm 1.1$. Similar anisotropy of Li$^+$ diffusion has been reported for the in-plane anisotropic GeP during electrochemical lithium intercalation with a ratio of $D_a/D_b \approx 25$.[21]

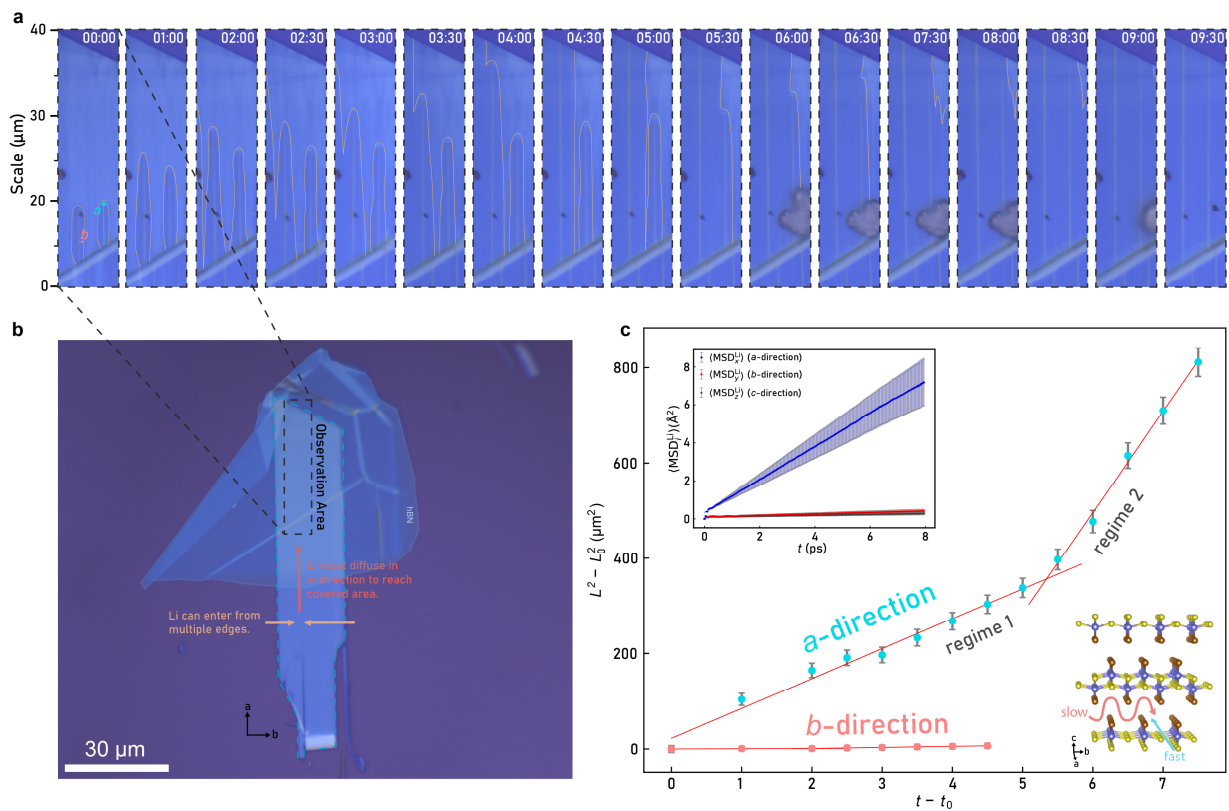

**Figure 4.** (a) Progression of Li$^+$ diffusion within the covered area of a partially hBN encapsulated sheet of CrSBr. A yellow line was added to easily distinguish the border between CrSBr and Li-CrSBr. To indicate the positions where the diffusion length is tracked, two arrows are added to the first panel for either direction, marked *a* and *b*. The gray particle seen between 06:00 and 08:00 is a lithium salt, which briefly deposited on the structure and was later washed away. (b) Optical micrograph of the whole heterostructure. The upper part is covered by hBN, the lower part is uncovered. During the intercalation, lithium can only enter the sheet from the exposed area and then diffuse into the covered part along the *a*-direction, until it reaches the observation area marked by the dashed line. (c) Plot of the diffusion progress as the relative increase of the square of diffusion length ($L^2 - L_0^2$) vs. $t - t_0$. The structure of CrSBr is shown on the bottom-right with the two diffusion directions marked by arrows. The inset in the top-left shows the computed direction-resolved lithium mean squared displacements at 600 K for comparison.

To further clarify the diffusion mechanism, we perform *NVT ab initio* molecular dynamics (AIMD) simulations (~50 ps) at four temperatures ranging from 600 to 1200 K on a Li-intercalated periodic CrSBr supercell containing 80 atoms, corresponding to 11% lithium intercalation (see Supplementary Information Section 6). In the supercell, the *x*, *y*, and *z* parameters correspond to *a*, *b*, and *c* lattice axis of CrSBr, respectively.

To quantify the diffusion anisotropy, we compute direction-resolved lithium mean squared displacements (MSD) from the Li-ion trajectories (inset in Figure 4c and Figure S16i-l), clearly revealing the dominance of the *x*-component. Importantly, the non-zero *z*-component does not imply that lithium ions are diffusing through the sheets in the *z*-direction. Instead, it corresponds to vertical ion movements required to access neighboring *x*-direction diffusion channels, as depicted schematically in the structural inset in Figure 4c and reported in the simulated Li-ion trajectories in Figure S15b (see also videos 1-6 in the Supplementary Information). Using the time-averaged Einstein relation (see Supplementary Information Section 6 and Refs.[22–26]), we estimate the total (Figure S17) and direction-resolved (Figure S18) self-diffusion coefficients at the four temperatures, obtaining a theoretical $D_x/D_y$ ratio of 20.7 ± 7.2 at 600 K, which is in -good agreement with the experimental value. The large uncertainty in the theoretical $D_x/D_y$ ratio arises from the very small value of $D_y$, which leads to a significant statistical error (see Figure S18). Above 600 K, $D_x/D_y$ ratio decreases significantly to 6.5 ± 3.3 at 800 K, followed by a more gradual reduction at higher temperatures, suggesting that the $y$-channels begin to open above 600 K. This trend is also evident in the Arrhenius plot for $Li^+$ diffusion (Figure S19). The more pronounced change in the $D_x/D_y$ ratio between 600 and 800 K, compared to the smaller variation between 800 and 1200 K, may indicate the presence of two distinct diffusion regimes, with a discontinuous transition occurring between 600 and 800 K, and a smoother evolution of anisotropy at temperatures below and above this range.

**Effect of Intercalation on Conductivity of Exfoliated Sheets**

A previous study on lithium intercalation of bulk CrSBr crystals found a significant increase in conductivity along the *a*-directions, while grain boundary formation impeded measurements along the *b*-direction.[17] We fabricate a six-terminal device to study the direction-dependent conductivity of exfoliated few layer sheets before and after intercalation. Furthermore, we investigate how the properties change over time when the device is exposed to air. The optical micrograph in Figure 5a shows a six-terminal device before and after intercalation as well as after 30 h of measurement. No damage to the flake is visible after intercalation but we can conclude it was fully intercalated from the Raman measurements in Figure 5c taken in the middle of the device. The contact configuration allows us to simultaneously study the effect of intercalation in either crystal direction on the same device; hence, we measure *I-V* sweeps along the two in-plane directions in an alternating fashion.

Figure 5b shows the temporal evolution of the current measured at a bias of 1.0 V along *a* and *b*-directions. Initially, the current is constant, around 0.6 µA for pristine CrSBr in both directions. Upon intercalation, the measurement is repeated in air. We immediately observe a sharp increase in conductivity to around 0.9 µA in the *a*-direction, in agreement with a previous reports[17]. The conductivity then drops over time and approaches a constant value of 0.4 µA, which is below the pristine value. Along the *b*-direction, on the other hand, the conductivity never exceeds the value before intercalation. Instead, it drops slightly at first and then follows a similar trend of steady decrease as observed in the *a*-direction, eventually reaching a constant value of 0.3 µA. Time-dependent *I-V* curves for each direction are shown in Figure 5d-e. In both directions, we observe a nonlinearity right after intercalation. Especially in the *b*-direction this likely contributes to the initial drop in the current vs. time plot. While improving slightly for the *b*-direction during the first few hours of the experiment, a nonlinearity in both directions persists and could be the reason why the current reaches a constant value below the pristine material.

We propose that two opposing effects are at play during the initial part of the measurement. Under perfect conditions, Li-CrSBr has an increased conductivity due to additional charge carriers from lithium anthracenide donating electrons to CrSBr during the intercalation and raising the Fermi-level. Nevertheless, at the same time the process of intercalation can worsen the interface between the gold contact and the CrSBr and introduce additional contact resistance or even the formation of a Schottky barrier leading to the nonlinearity in the *I-V* curves and reduced current observed in the measurement. Additionally, especially in the *b*-direction, the intercalation can introduce microtears and new grain boundaries, which explains the observed difference in *a* and *b*-direction.

The general decrease in conductivity over time shows that Li-CrSBr is not air stable and will gradually lose its lithium, most likely through reaction with moisture in the air. The third Raman spectrum (30 h Air) in Figure 5c confirms that after 30 h in air the intercalation has been almost completely reversed, which is indicated by the reappearance of pristine Raman modes with only small shoulders remaining. This deintercalation process appears to happen uniformly as both curves follow a similar trajectory for most of the measurement. Further measurements (Supplementary Information Section 7) confirm this trend of gradual deintercalation upon exposure to air.

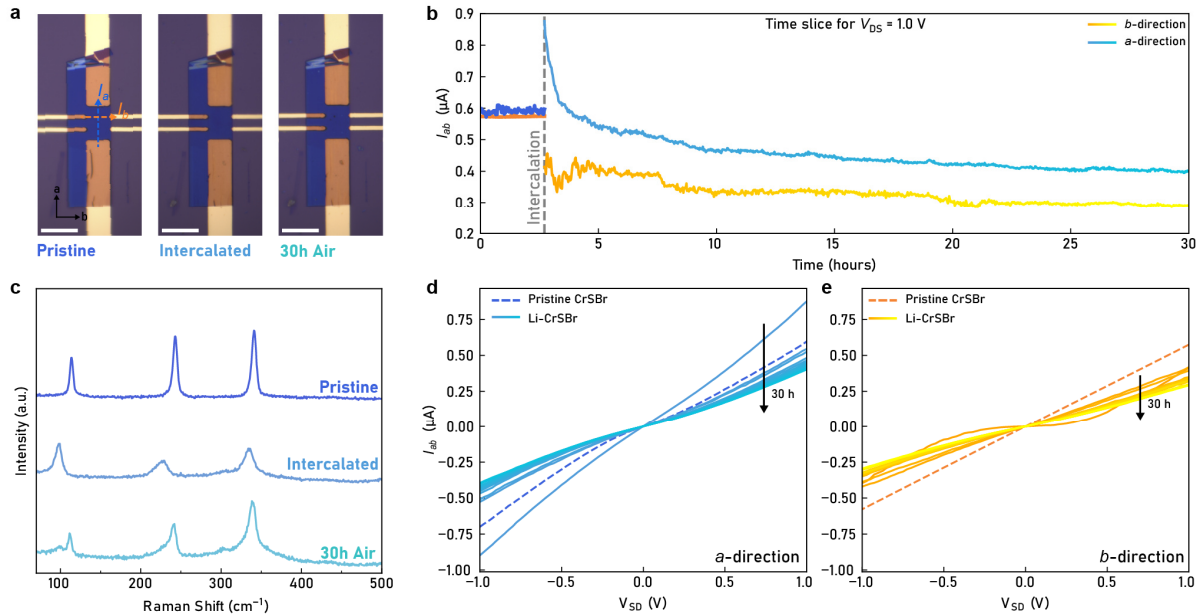

**Figure 5.** (a) Optical Micrograph of CrSBr devices before intercalation, after intercalation, and after 30 h of *I-V* sweeps. (b) Time dependence of current before and after intercalation. Data was extracted from the *I-V* sweeps for a bias of 1.0 V. (c) Raman spectra of the pristine and intercalated device and after 30 h of measurements (taken in the middle of the device). (d-e) *I-V* sweeps in the *a* and *b*-direction at different times over a period of 30 h.

## Summary

Our work demonstrates that lithium can be intercalated into exfoliated CrSBr. This process is highly anisotropic, which is in good agreement with atomistic simulations. We tracked the intercalation dynamics by leveraging Raman spectroscopy, optical contrast, and electrical conductivity measurements and revealed that lithium intercalation serves as an effective tool of tuning the electronic properties of CrSBr. The ability to intercalate lithium in a controlled manner using partial hBN encapsulation establishes a robust platform for investigating alkali metal intercalation into layered 2D materials, offering new opportunities for tailoring the properties of van der Waals materials for future applications in electronics, energy storage, and quantum materials research.

## Methods

### Exfoliated CrSBr sample fabrication

CrSBr bulk crystals were synthesized by chemical-vapor transport (CVT) as described in Ref. [8] We employed a conventional Scotch tape method to exfoliate bulk CrSBr and transfer exfoliated material on top of $Si/SiO_2$ substrates to conduct optical microscopy and Raman spectroscopy studies. All studies described in this paper were conducted on the $SiO_2$ (300 nm)/$p^+$ Si substrates. To meet the requirements of the secondary ion mass spectrometry method, we prepared gold-coated (250 nm) $Si/SiO_2$ substrates by metal evaporation and transferred exfoliated flakes simply by Scotch tape method. For electronic characterization of CrSBr devices, the CrSBr and hBN was exfoliated on polydimethylsiloxane (PDMS) and transferred via the common PDMS dry transfer method. For this the glass slide was loaded into the prober arm of the transfer setup and the flake was transferred on top of a $Si/SiO_2$ substrate with prefabricated gold electrodes. The stage was heated to 60 °C, maintained at that temperature for 30 s, and then the glass slide was slowly lifted. For hBN/CrSBr heterostructures a slightly different pickup and dry transfer method was employed. A thick PDMS sheet was cut into the shape of a pyramid and covered with a thin layer of polycarbonate (PC) which was obtained by drop casting a 7 wt.% solution of PC in chloroform between two glass slides and letting the chloroform evaporate on air. The PDMS stamp was loaded into the prober arm of the transfer setup and used to pick up exfoliated hBN from PDMS at 90 °C. Subsequently the CrSBr flake was picked up through van der Waals interactions with the hBN sheet on the PC stamp. Finally, the heterostructure was released on an $SiO_2/Si$ substrate with prefabricated gold electrodes at 185 °C. The remaining polymer was dissolved by submerging the device in chloroform for 2 h at room temperature. All transfers were carried out in an argon atmosphere to avoid sample degradation. All devices were annealed to improve their electrical performance (5 h, 300 °C, $10^{-6}$ mbar).

### Lithium-solvated electron solution

Lithium anthracenide stock solution was prepared in following way: Anthracene (77.40 mg, 0.434 mmol) is first dissolved in dry THF (4.55 ml) followed by the reaction with metallic lithium (2.00 mg, 0.288 mmol) under an inert argon atmosphere. The brilliant blue color of the resulting solution immediately arises from the presence of radical anions. Lithium anthracenide stock solution molarity is 32 mM. Note that we would like to keep anthracene in excess to make sure the limiting factor is not the availability of anthracene. The dry solvent is ultimately necessary since even traces of water or oxygen will destroy the radical reagent. Under careful observation, we exposed the reaction mixture to gentle manual shaking without stirring until the reaction was complete manifesting a deep blue color and no traces of metallic lithium on the surface. Therein, the prepared stock solution was kept at a cooling temperature of −30 °C. To make the solution suitable for intercalation it needs to be diluted with dry THF by a factor of ~100 to reach a concentration of approximately 0.32 mM. The exact dilution factor may vary *e.g.* due to residual moisture in the reagents, so we suggest monitoring the color of the solution during the dilution process. The point where the bluish hue has just vanished and the solution is colorless usually corresponds to a concentration well suited for intercalation.

**Characterization Methods**

Ambient condition current measurements were carried out at room temperature with a six channel Keysight B1500 Precision Source/Measure Unit (SMU) in an INSTEC 4-probe measurement setup. Raman spectral measurements and mapping were conducted with a WITec Confocal Raman Microscope (WITec alpha300 R, Ulm, Germany), equipped with a 532 nm laser and a spectrometer with a thermoelectrically cooled CCD camera sensor. The measurements were performed in an argon-filled glovebox at room temperature with a 100× objective and a laser power of less than 2.5 mW to avoid sample degradation. Characterization by Atomic Force Microscopy (AFM) was performed on a NanoMagnetics ezAFM compact atomic force microscopy system in dynamic mode (tapping mode) inside an argon-filled glovebox. Temperature dependent conductivity measurements were performed using the Electrical Transport Option (ETO) of the Physical Property Measurement System (PPMS) instrument manufactured by Quantum Design.

**DFT Phonon Calculations**

To verify our Raman shifts and the structure, we optimized the CrSBr and CrSBr-Li unit cell structure, including the atom positions in the QUANTUM ESPRESSO software package[27,28], rev 7.1. The unit cell of CrSBr consisted of two Cr, two S and two Br atoms; Li-CrSBr contained also one lithium atom (a similar structure was also proposed for FeOCl-Li[10]). The PBE functional was used for all calculations with the projector-augmented wave method (PAW) obtained from the PSEUDOJO library[29] with a wavefunction kinetic energy cutoff $E_{cut}$ = 50 Ry and a charge density energy cutoff $E_{cut}$ = 500 Ry. The convergence criteria for the cell optimization were: energy <$10^{-6}$ Ry, force <$10^{-4}$ Ry/Bohr. The required SFC convergence was set to <$10^{-15}$ Ry. Requested k-spacing 0.15 Å$^{-1}$, leading to 12 × 9 × 6 mesh, was used. Phonon modes were calculated at the Γ point.

**AIMD Simulations**

We calculated Li$^+$ diffusion in CrSBr by performing *ab initio* Born-Oppenheimer molecular dynamics simulations using a Li-intercalated 80-atom supercell (Li$_8$Cr$_{24}$S$_{24}$Br$_{24}$) with 11% lithium intercalation under periodic boundary conditions. Details of the fully relaxed (atomic positions and cell vectors) Li-CrSBr cell are reported in the Supplementary Information Section 6. The electronic degrees of freedom were dealt with using DFT within the plane-wave pseudopotential formalism ($E_{cut}$ = 40 Ry, GBRV ultrasoft pseudopotentials,[30] Brillouin zone integration at the Γ point, and PBEsol exchange-correlation functional[31]). The ions were propagated for 50 ps using an *NVT* ensemble with a stochastic velocity rescaling thermostat.[32] The simulations were performed at temperatures of 600, 800, 1000, and 1200 K using the QUANTUM ESPRESSO software package software package[27,28], rev 7.1. The effect of lithium intercalation on the volume was considered by conducting the *NVT* runs on supercells preliminarily equilibrated through *NPT* ensemble simulations for ~10 ps. The high temperature regime was chosen due to the low mobility of the Li atoms. Total energies and temperatures during these simulations are reported in the Supplementary Information. Diffusion coefficients were extracted from the Li trajectories according to the Einstein relation.[22] Details on the method to extract diffusion coefficients and statistical variances are given in the Supplementary Information Section 6 and in Ref.[25].

**TOF-SIMS**

The Gallium ion source based FIB-SEM and TOF-SIMS system used was composed of the TESCAN LYRA instrument platform on which was mounted the TOF-SIMS analyzer (TOFWERK AG) developed collaboratively by EMPA (Swiss Federal Laboratories for Materials Science and Technology, Thun) and TESCAN s.r.o. (Brno, CZ). The acquisition was carried out at the ion beam energy of 20 keV and current of 260 pA. Mass count from the scanning area was recorder in 200×200 pixels maps with binning factor of 2×2, which eventually resulted in square pixels with size in x- and y- dimensional (top side of the crystal) of 200 nm and 300 nm for scanning view field of 20×20 μm$^2$ and 30×30 μm$^2$, respectively. The pixel size in the z-direction is 0.1 nm per frame.

## Acknowledgments

Z.S. was supported by European Union's Horizon Europe research and innovation programme under grant agreement ID 101135196. This work was supported by project LUAUS23049 from Ministry of Education Youth and Sports (MEYS) and Marie Curie Sklodowska ITN network "2-Exciting" (Grant No. 956813). K.M. acknowledges the grant of specific university research A2 FCHT 2024 055. The authors acknowledge the assistance provided by the Advanced Multiscale Materials for Key Enabling Technologies project, supported by the Ministry of Education, Youth, and Sports of the Czech Republic (project No. CZ.02.01.01/00/22_008/0004558 co-funded by the European Union). M.V. further acknowledges the Czech Science Foundation (GACR No. 23-08083M) for financial support Computational resources were provided by the e-INFRA CZ project (ID:90254), supported by the Ministry of Education, Youth and Sports of the Czech Republic. F.D. gratefully acknowledges funding from the Emmy Noether Program (Project-ID 534078167). Low temperature experiments were performed in MGML (mgml.eu) supported within the program of Czech Research Infrastructures (project no. LM2023065). The AIMD simulations work was supported by a grant from the Swiss National Supercomputing Centre (CSCS) under project ID mr33 (Eiger).

## Author Contributions

J.S., M.V., and Z.S. initiated and supervised the project. K.M. and Z.S. grew the CrSBr crystals. K.M. and A.S. fabricated heterostructures and performed intercalations as well as Raman measurements. A.S. performed transport measurements. K.M. and A.S. analyzed the data with input from F.D., J.S., M.V. and D.S. G.M., N.M., and G.-M. R. performed molecular dynamic calculation and analyzed the data. J.S. performed DFT calculations. P.L. performed low-temperature transport measurements. M.V. conducted TOF-SIMS analysis. AFM measurements were done by B.R. K.M. and A.S. wrote the manuscript with input from all authors.

# Supplementary Information:

Lithium Intercalation in the Anisotropic van der Waals Magnetic Semiconductor CrSBr

*Kseniia Mosina[1†], Aljoscha Söll[1†], Jiri Sturala[1*], Martin Veselý[2*], Petr Levinský[3], Florian Dirnberger[4], Giuliana Materzanini[5], Nicola Marzari[6], Gian-Marco Rignanese[5,7], Borna Radatović[1], David Sedmidubsky[1], and Zdeněk Sofer[1*]*

[1] Department of Inorganic Chemistry, University of Chemistry and Technology Prague, Technická 5, 166 28, Prague 6, Czech Republic

[2] Department of Organic Technology, University of Chemistry and Technology Prague, Technická 5, 166 28, Prague 6, Czech Republic

[3] Institute of Physics of the Czech Academy of Sciences, Cukrovarnická 10, 162 00, Prague 6, Czech Republic

[4] Department of Physics, TUM School of Natural Sciences, Technical University of Munich, Garching, Germany

[5] Institute of Condensed Matter and Nanosciences (IMCN), Université catholique de Louvain, Louvain-la-Neuve 1348, Belgium

[6] Theory and Simulation of Materials (THEOS), and National Centre for Computational Design and Discovery of Novel Materials (MARVEL), École Polytechnique Fédérale de Lausanne (EPFL), CH-1015 Lausanne, Switzerland

[7] Wel Research Institute, Wavre, Belgium

† These authors contributed equally.

**Keywords**: lithium, van der Waals materials, intercalation, anisotropy.

# Supplementary Section 1. Optical Contrast Techniques to Estimate the Thickness of CrSBr.

Optical contrast is a well-established method for estimating the thickness of two-dimensional materials in white-light microscopy.[1–3] This technique, commonly applied to graphene, h-BN, and TMDCs flakes on $SiO_2$/Si substrate allows for thickness estimation as a function of their color.[4,5] Figure S1 displays the optical images of mechanically exfoliated bulk crystals with corresponding profile data details. A visible color shift from violet-blue to yellow and grayish-pink indicates changes in flake thickness, as reflected in the developed color scale.

In accordance with the group theory, Raman scattering spectra of CrSBr at 532 nm excitation exhibit the three active phonon modes ($A_g$, $B_{2g}$, $B_{3g}$) observable at room temperature. Among them, the main optical phonons are those of $A_g$ symmetry. Atomic displacements of the $A_g^1, A_g^2, A_g^3$ modes correspond to out-of-plane vibrations of chromium, sulfur, bromine atoms, respectively. Considering the anisotropic nature of CrSBr, we co-polarized the excitation along the crystalline $b$ axis. Thus, we observe out-of-plane $A_g^1$, $A_g^2$, and $A_g^3$ modes with the frequency of 114 cm$^{-1}$, 244 cm$^{-1}$, 342 cm$^{-1}$ for mechanically exfoliated CrSBr (Figure S1d). Remarkably, the Raman spectra do not differentiate the thickness of all selected flakes with the number of layers N > 10.

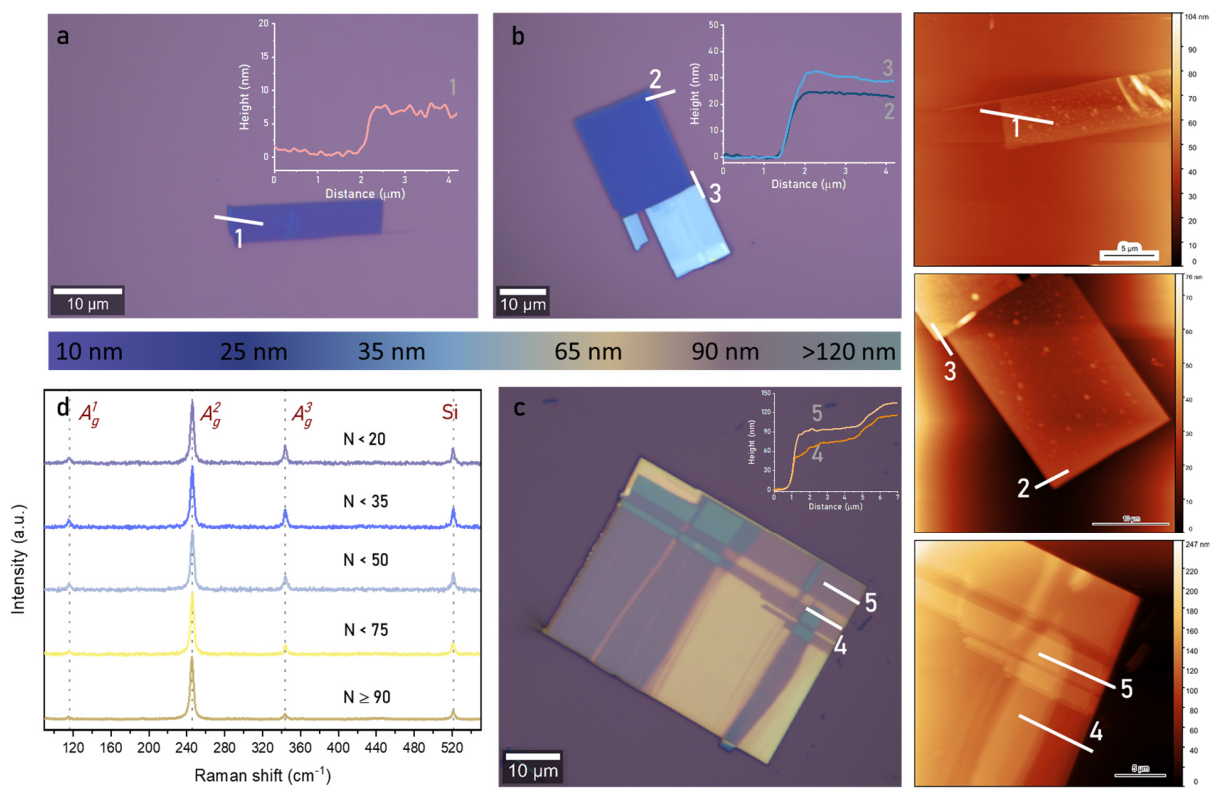

**Figure S1.** Characterization of exfoliated CrSBr. (a) Optical micrograph of ca. 8 nm thick flake exhibiting a violet blue color. (b) Ca. 25 nm and ca. 35 nm thick flakes exhibiting a bright blue and sky-blue color, respectively. (c) Multilayered flake with increasing thickness values from ca. 65 nm to more than ca. 90

nm manifested by different colors from yellow to grayish pink. (d) Raman spectra of selected flakes of different thickness. The color scale for rough thickness estimation is presented in the middle. Corresponding AFM micrographs and line scan profile 1 of (a) ca. 8 nm thick flake, profile 2 and 3 of (b) ca. 20 nm and ca. 30 nm thick flakes, respectively, and profiles 4 and 5 of (c) multilayered flake with a thickness of ca. 65 nm and 90 nm, respectively.

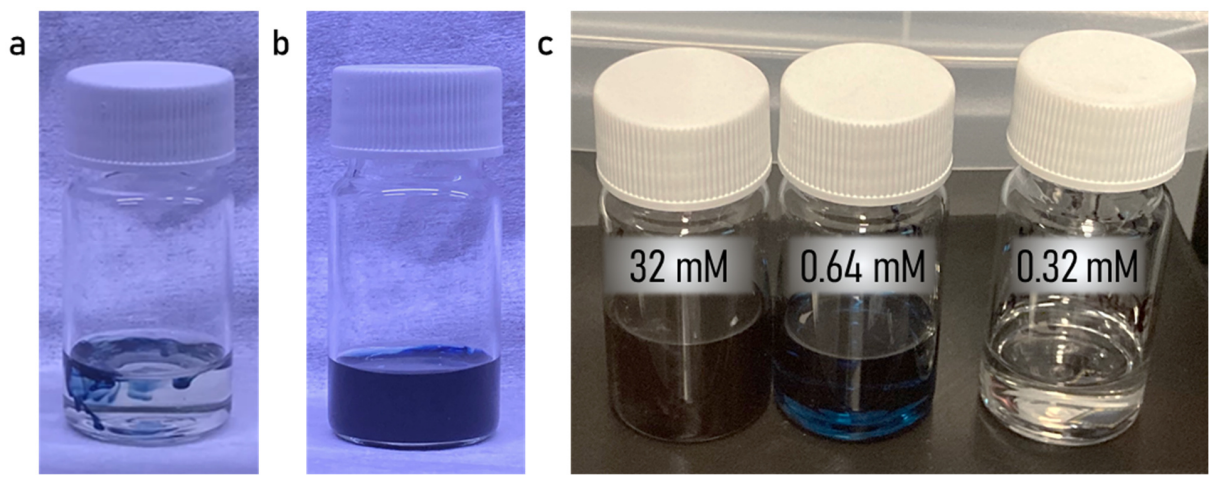

**Figure S2.** Optical micrographs of the solvated electron solution. (a) Lithium anthracenide in THF at the onset of the reaction, when thin plates of metallic lithium are added to the mixture and (b) after stabilization, when the reaction is completed and all lithium dissolved, the solution turns a saturated deep blue. The stock solution (32 mM lithium anthracenide) appears dark blue, as shown in micrographs (b) and (c). In micrograph (c), the diluted stock solutions (0.64 mM and 0.32 mM lithium anthracenide) exhibit a brilliant blue and a transparent appearance, respectively.

## Supplementary Section 2. Secondary-ion mass spectrometry for the study of lithium intercalation

Since lithium compounds are challenging to analyze utilizing conventional EDX spectroscopy techniques due to the low characteristic X-ray signal of lithium, even with a windowless detector, we employed time-of-flight secondary-ion mass spectrometry (TOF-SIMS) technique. This method relies on sputtering the sample's surface with a beam of primary ions and detecting the mass of secondary ions formed during the sputtering process[6], confirming the successful intercalation of Li into CrSBr structure highlighting the presence of lithium within the depth of material.

Secondary ion intensity can be expressed by a value that accounts for the complex matrix effects and sputtering conditions:

$$I_s = I_p Y R^{\mp} c^{surf} T \quad (S1)$$

where $I_p$ is the primary ion current, $Y$ is the sputtering yield, $R^{\mp}$ is the ionization probability, $c^{surf}$ is the fractional concentration of the element in the surface, and $T$ is a mass spectrometer transmission. However, applying identical acquisition conditions, matrix effect of the studied material and the same settings of mass spectrum (MS), we can assume a simplified equation

$$I_s = X(I_p, Y, R^{\mp}, T)\, c^{surf} \quad (S2)$$

where $X$ is combined constant, which is the same among all the measurements for every specific mass (note, that there is no correlation between various masses). Under these assumptions, we can determine relative change in concentration both in lateral sizes (top view) and in depth profile.

We prepare the sample of mechanically exfoliated CrSBr on a silicon substrate coated with a 250 nm-thick gold layer to avoid electron charge accumulation on the wafer, since the electrons recombine with ions and therefore, a significant loss of MS signal is observed. The wafers with mechanically exfoliated CrSBr were characterized using optical, Raman, and atomic force microscopy, then immersed in 0.32mM lithium anthracenide solution for 15 minutes to perform the intercalation. Figure S3 presents optical micrographs, which provide an overview of the pristine and intercalated material. Based on the observed color, we concluded that flakes 1 and 2 possess similar thicknesses, whereas flake 3 is significantly thicker. This observation is confirmed by AFM analysis. The thickness of flake 1 is approximately 15 nm, flake 2 is 20 nm, and flake 3 is 70 nm, with a thicker edge measuring ~120 nm (indicated by the red, green, and blue profiles in the optical image inset, respectively). Raman spectra presented in Figure S3a-c confirmed that complete intercalation was achieved for the thin flakes (1 and 2) exhibiting the low-energy vibrational mode $A_1^1$, while the thicker flake (3) was partially intercalated.

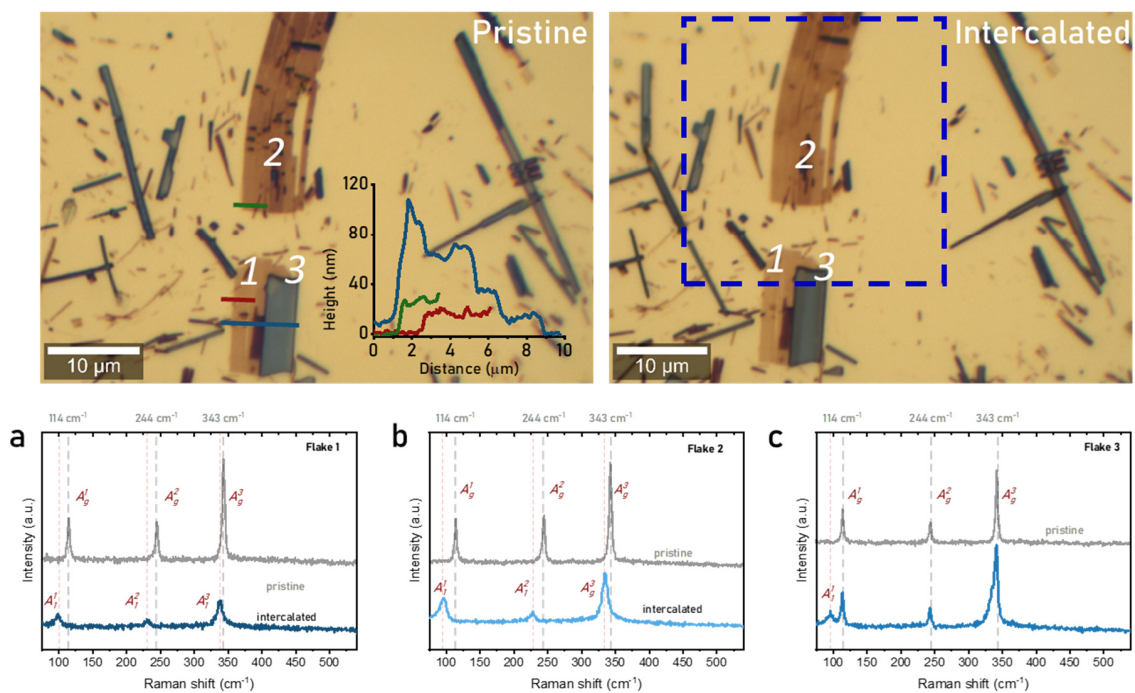

**Figure S3.** Optical image of pristine exfoliated CrSBr with AFM profiles of marked flakes 1-3 shown in the inset. Optical image of intercalated flakes with a marked by a blue dotted rectangle area chosen for SIMS analysis (30 × 30 μm). (a–c) Raman spectra of pristine and intercalated flakes 1-3.

Secondary ions masses generated from the intercalated sample were detected at a given polarity (positive and negative) while maintaining the same settings for the primary $Ga^+$ source, which ensured a consistent sputtering rate. Although lithium is easily ionized due to its high ionization probability resulting in a strong signal, the matrix effect poses significant challenges for quantifying lithium ions without suitable standards, such as implanted elements within the same matrix.[7] Consequently, both lithium (the element of interest) and chromium (the material element) intensities were observed in the positive ion elemental mapping, as the inherently measured signals under identical acquisition conditions served as a relative concentration of the specific element.

The schematics in Figure S4a illustrate the principles of SIMS analysis. First, we collected the signal from regions of 100 × 100 pixels within a field of view of 30 × 30 μm to fairly represent lithium distribution within the studied sample. Once primary ionization was initiated, Li-CrSBr was analyzed frame by frame, while the secondary ion masses generated from increasing sputtered depths were detected. This spectroscopy method is destructive, as sputtering etches both the material and the substrate at the same rate, as depicted in the SEM micrographs of the initial and analyzed samples in Figure S4b,e. The resulting positive secondary ion image maps (30 μm × 30 μm) in Figure 4c,d,f and g display the elemental distribution of Li and Cr across the area of interest. The color indicates the number of selected species emitted per pixel within the analyzed area, ensuring both Li and Cr are present. Additionally, the positive ion mass spectral data shows all isotopes of Cr and Li and indicates the signals of lithium *(m/z 7)* and chromium *(m/z 52)* with their intensities an order of magnitude higher than those of environmental sodium and potassium, as presented in Figure S4h.

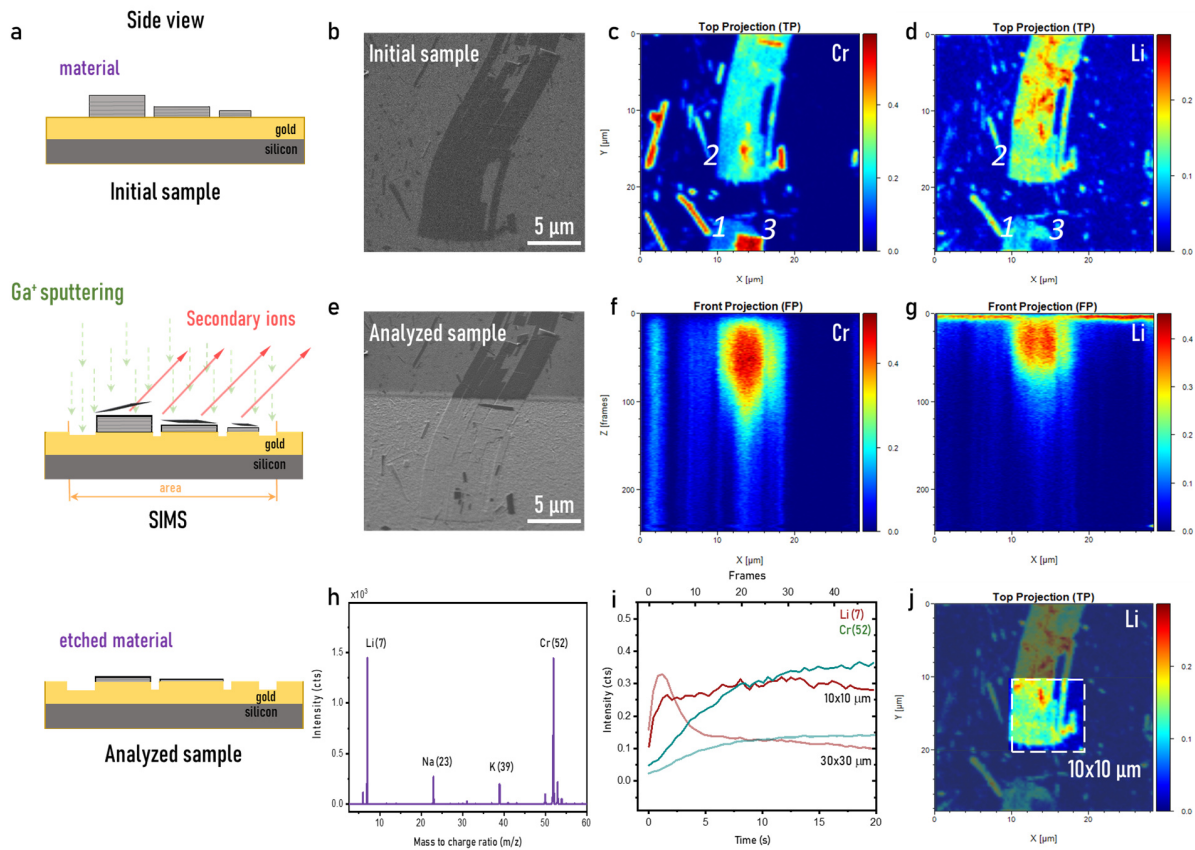

**Figure S4.** Secondary-ion mass spectrometry implementation. (a) Schematic illustration of SIMS principals. SEM micrographs of initial (b) and analyzed (e) sample taken before and after spectrometry, respectively. The secondary ion image maps of elemental chromium and lithium represented in top (c, d) and front (f, g) projections. (h) Secondary ions mass spectrum of Li-CrSBr collected from whole (30 × 30 μm) area. (i) Elemental depth profiles of Li (7) and Cr (52) collected from 30 × 30 μm and limited 10 × 10 μm areas of interest. (j) Limited to 10 × 10 μm area of interest.

Although lithium and chromium exhibit strong signals within the 100 frames, as illustrated in the front projection maps in Figure S4f,g, their intensities evolve distinctly as sputtering progresses. However, the view field of 30 × 30 μm spans not only the intercalated CrSBr flakes but also the wafer itself, which affects the resulting front projection maps. Thus, we limited the area of interest to 10 × 10 μm to ultimately represent the element depth analysis from the intercalated material. The considered area is presented in Figure S4j. The element depth analysis of the limited area seen in Figure S4i shows that elemental intensities of Li and Cr are in the same order of magnitude with an increase within the slight penetration, which evidence the ultimate presence of Li within the CrSBr crystalline structure. Notably, the lithium profile stabilizes shortly after sputtering begins, in contrast to the depth profile obtained from the 30 × 30 μm view field. After 200 frames, both Cr and Li signals are significantly weakened manifesting the etching and destruction of the investigated material. Upon completion of the scan, we quantified the penetration depth based on the intensity of the Cr signal and the known thickness of the investigated flake measured by AFM. Since the depth profile was collected from flake 2 with a known thickness of 20 nm, we conclude that 1 nm of material is etched within 10 sputtering frames.

By surveying a series of intercalation conditions, we find that a 0.32 mM lithium anthracenide solution (15 min immersion) leads to the Li/Cr secondary-ion ratio within the same order of magnitude. As expected, the highest lithium intensity was observed for the Li-CrSBr treated with 0.45 mM lithium anthracenide, however, the signal of Cr was lower by an order of magnitude compared to Li, while lower lithium anthracenide concentration (0.2 mM) resulted in incomplete intercalation.

To investigate the influence of 0.32 mM lithium anthracenide on the intensity of secondary ions masses generated from the intercalated sample, we further acquired the SIMS signal from the sample treated with an abundance and deficiency of lithium ions. Figure S5 illustrates the variations in the intensities of the element of interest (Li) and the material element (Cr) as a function of the lithium anthracenide concentration used for intercalation. The data points were extracted from the elemental depth profiles of 13 samples after 10 seconds of sputtering, with the considered area of 10 × 10 μm, covering exclusively the surface of the material.

The analysis of the samples treated with 0.45 mM lithium anthracenide revealed the highest lithium intensity, with the intensities of Li (7) and Cr (52) differing by 7.23 ± 2.26 cts. Conversely, the sample treated with 0.32 mM lithium anthracenide exhibited the smallest intensity ratio between Li (7) and Cr (52), calculated as 0.57 ± 0.24 cts, and represented by overlapping data points in Figure S5. A deficiency of lithium ions, achieved by treatment with 0.2 mM lithium anthracenide, resulted in a weaker lithium signal, with an intensity ratio between Li (7) and Cr (52) of 0.33 ± 0.24 cts. Collectively, these findings establish a trend between the generated secondary ion masses and the concentration of lithium anthracenide used, confirming the Raman study findings claiming 0.32 mM lithium anthracenide is the "optimum" concentration for 15 minutes of CrSBr intercalation.

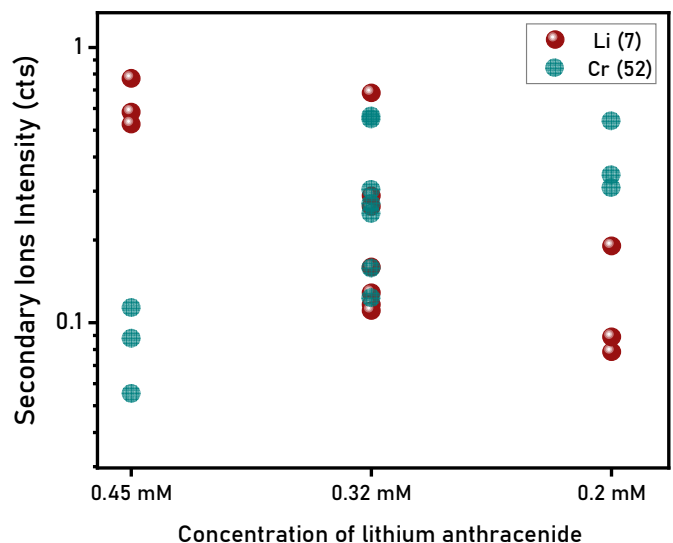

**Figure S5.** The variations in the intensities of the element of interest (Li) and the material element (Cr) as a function of the lithium anthracenide concentration.

**Supplementary Section 3.** Lithium Intercalation into Exfoliated CrSBr: Concentration Experiment.

Figure S6 presents optical micrographs and Raman spectra of exfoliated CrSBr flakes before and after lithium intercalation. On each wafer, eight flakes with varying colors, ranging from violet-blue (thin) to grayish-yellow (thick), were selected to represent different thicknesses, labeled 1 through 8 accordingly.

To assess whether anthracene in THF affects the CrSBr structure, we first conducted a control experiment using a blank solution (no lithium, no solvated electrons). As shown in Figure S6a, the optical appearance of the flakes remained unchanged, and Raman spectra (Figure S6f) showed no shifts in the $A_g^1, A_g^2$, and $A_g^3$ modes (114, 244, and 342 cm$^{-1}$, respectively). All peaks exhibited clean Lorentzian line shapes, from which we extracted peak positions and FWHM values (plotted as error bars).

We started with 0.2 mM lithium anthracenide (32 mM stock diluted by factor of 150) for CrSBr intercalation. At this concentration, the solution appeared notably transparent. As seen in Figure S6b, thicker flakes darkened slightly around the edges. Figure S6g illustrates the frequency positions of the $A_g^1, A_g^2, A_g^3$ modes, where solid-colored spheres are original peak positions, and semi-transparent spheres depict the data from Li-CrSBr. Intercalated flakes 1-4 with a thickness of ca. 30 nm exhibit $A_1^1, A_1^2, A_1^3$ modes at 95 cm$^{-1}$, 225 cm$^{-1}$, and 334 cm$^{-1}$, respectively, indicated by the dashed line in Figure S6g. Additionally, the FWHM values of the intercalated flakes are broadened, which shown by the error bars.

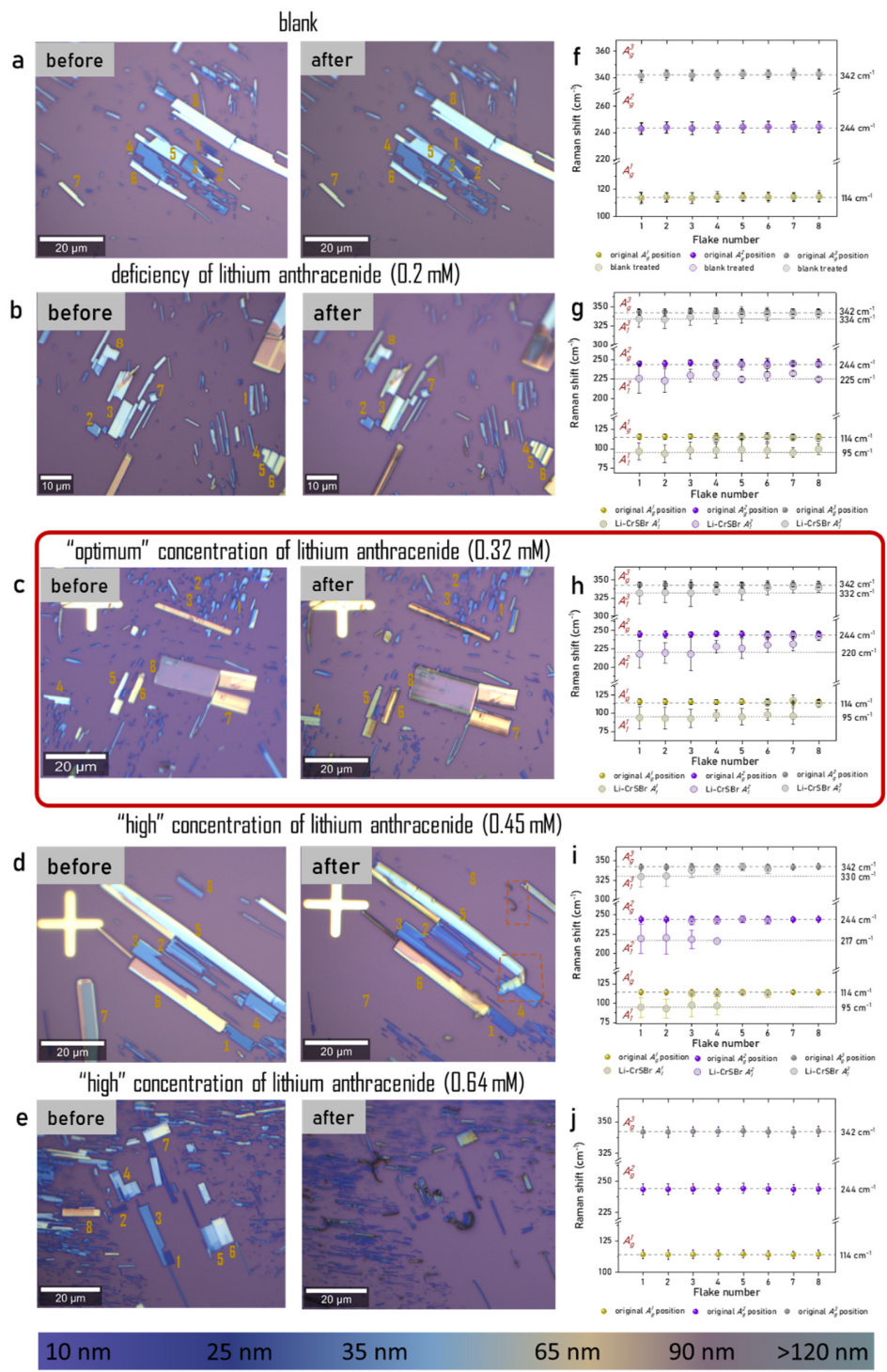

**Figure S6.** Concentration series experiment: optical micrographs of the exfoliated flakes before and after lithium anthracenide treatment with (a) blank with no lithium and varying concentration of (b) 0.2 mM, (c) 0.32 mM, (d) 0.45 mM, and (e) 0.64 mM lithium anthracenide. (f–j) Raman mode positions as a function of flake number marked on the corresponding optical micrographs.

To gain further insight, we analyzed the intercalation using a 0.32 mM lithium anthracenide solution (stock diluted by a factor of 100), which remarkably remained transparent. Optical images presented in Figure S6c reveal minimal changes in thin flakes (1–3), with a mild darkening, while thicker flakes (6–8) displayed increased contrast at the edges. Importantly, all flakes remain present, with no external morphological changes visible. Raman analysis of Li-CrSBr flakes 1–3 with a thickness of ca. 30 nm in Figure S6h confirmed $A_1^1, A_1^2, A_1^3$ modes at frequencies of approximately 95 cm$^{-1}$, 219 cm$^{-1}$, and 331 cm$^{-1}$, respectively. The FWHM of the Li-CrSBr flakes are broadened, with FWHM of 14 cm$^{-1}$ for Li-CrSBr and FWHM of 4 cm$^{-1}$ for CrSBr, as shown by the error bars in Figure S6h.

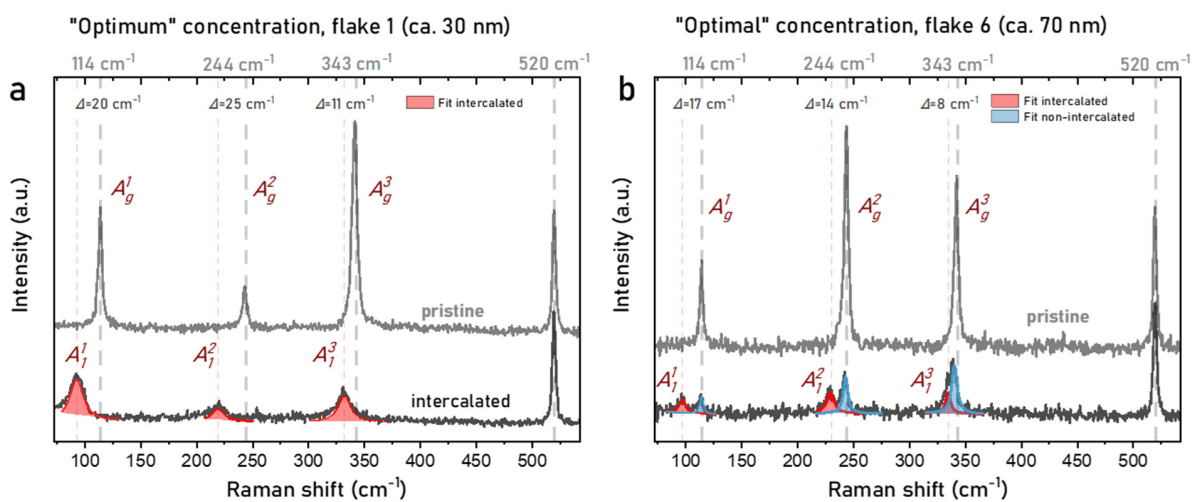

**Figure S7.** Raman spectra of exfoliated CrSBr before (pristine), and after intercalation of ca. 30 nm thick flake 1 and ca. 70 nm thick flake 6. Silicon peak at 520 cm$^{-1}$ originates from the substrate.

Figure S7a shows the Raman spectra of a representative fully intercalated flake with deconvoluted $A_1^1, A_1^2, A_1^3$ modes, picturing a shift in the $A_g$ modes frequency positions together with a notable intensity decrease. The Raman spectra of a partially intercalated flake with a thickness of ca. 70 nm, presented in Figure S7b, reveal two distinct sets of phonon atomic displacement modes, indicating the presence of both intercalated and non-intercalated material within the illuminated area. Similar phenomena were noticed for the 0.2 mM lithium anthracenide, where we observed partial intercalation of the flakes thicker than 30 nm. In contrast, flake 8 remained unaffected (Figure S6h), with no $A_1$ modes detected.

Intercalation with 0.45 mM lithium anthracenide solution (stock diluted by factor of 75) demonstrated the excess of lithium, reflected at first in the distinctions in the optical images taken before and after the experiment (Figure S6d). Upon intercalation, flakes 1–4 turned bright blue, flakes 5–6 showed edge darkening, and flakes 7–8 were no longer present. Notably, the morphology of the flakes marked by dashed rectangles was altered: the thin flakes appear distorted, and part of the thicker flake is bent over. Raman spectra of flakes 1 and 2 reveal that $A_1^1, A_1^2, A_1^3$ modes consistently appear at frequencies of approximately 95 cm$^{-1}$, 217 cm$^{-1}$, and 330 cm$^{-1}$, respectively, corresponding to Raman shifts of 19 cm$^{-1}$, 27 cm$^{-1}$, and 12 cm$^{-1}$ from original values. Furthermore, the Li-CrSBr FWHM values resulted in even greater FWHM broadening of 15 cm$^{-1}$ in comparison with FWHM of 4 cm$^{-1}$ for CrSBr, shown by error bars in Figure S6i. Moreover, Raman spectra of flakes 3 and 4 show two phonon atomic displacement modes consistent with previously reported values for $A_g$ and $A_1$ displacements. This behavior differs from thicker flakes 5 and 6, which remain structurally preserved, showing the frequency positions of the $A_g^1, A_g^2, A_g^3$ modes on the same positions.

To further investigate the effect of excess lithium, we used 0.64 mM lithium anthracenide in THF (stock diluted by a factor of 50). We note that solution exhibits a bright blue color. This reagent concentration significantly affects the appearance and crystalline structure of exfoliated CrSBr, as evidenced by the optical images captured before and after lithiation (Figure S6e). The intercalation with an excess of lithium causes irreversible changes in the crystalline structure of CrSBr flakes, thus, $A_g$ vibrational modes are not clearly observed in the Raman spectra of treated flakes (Figures S6j and S8).

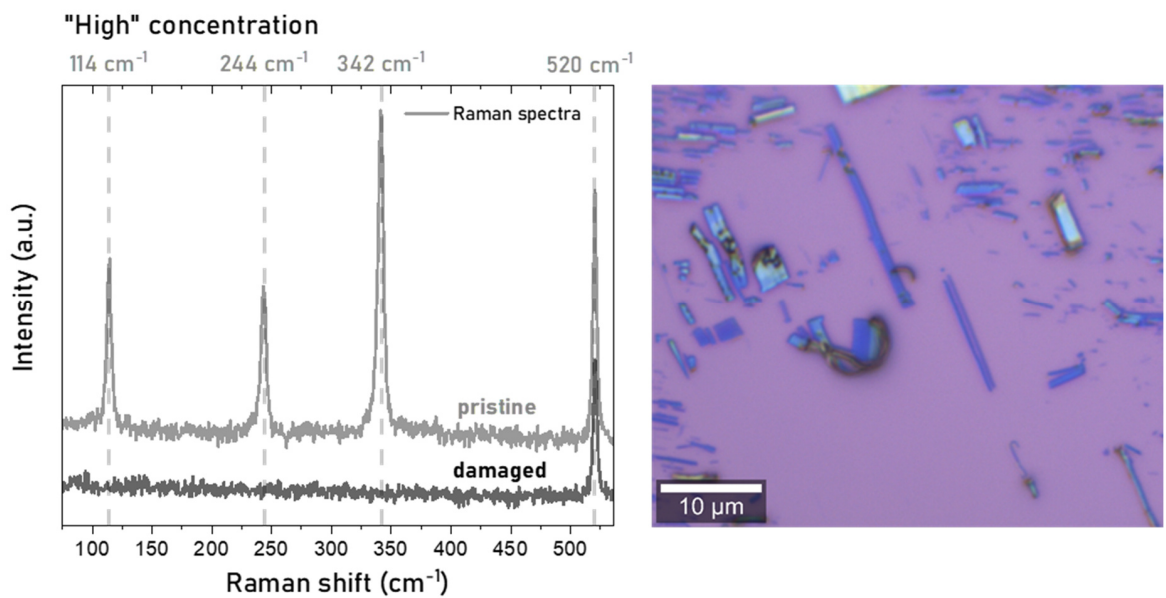

**Figure S8.** Optical micrograph after intercalation and Raman spectra of exfoliated CrSBr after the process (damaged). Due to the excess of the intercalant, the investigated flake loses its crystallinity, resulting in the suppressed frequency vibrations of Ag modes. Silicon peak at 520 cm$^{-1}$ originates from the substrate.

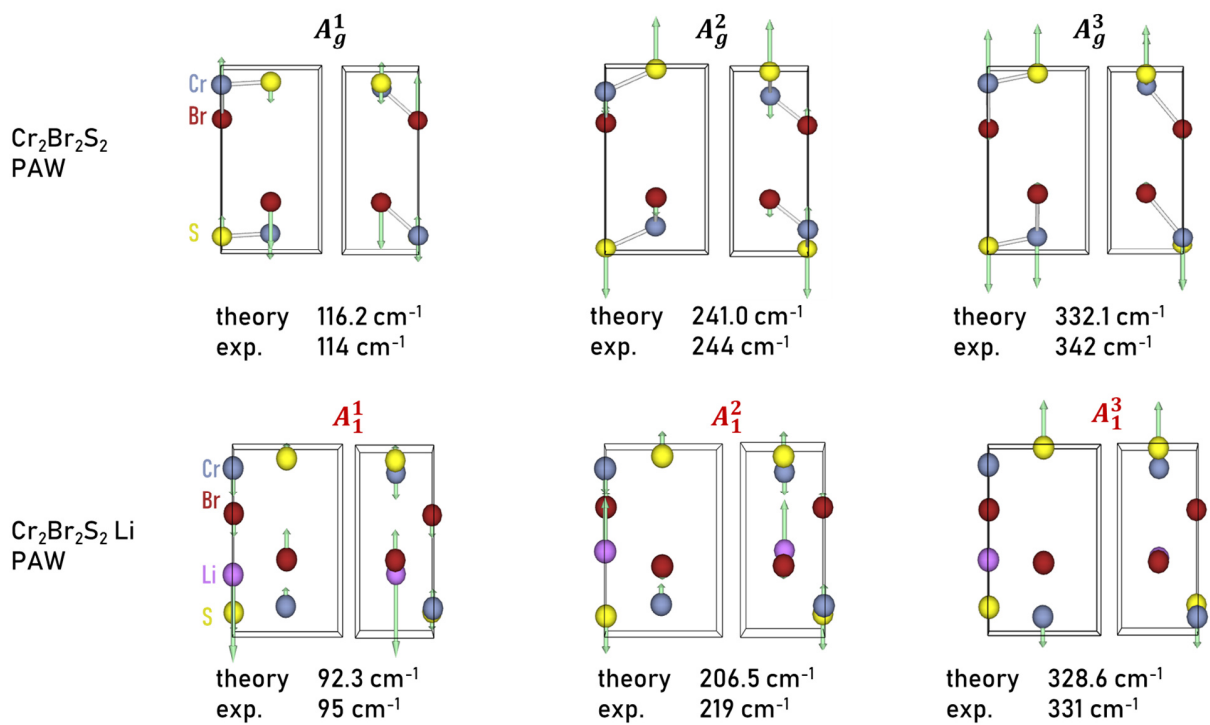

**Figure S9**. DFT calculated optical phonon modes for pristine CrSBr and $Li_{0.5}$-CrSBr.

## Supplementary Section 4. Effects of partial and full hBN encapsulation.

In the process of studying the intercalation dynamics in Li-CrSBr we fabricated several flakes of CrSBr either fully or partially intercalated with hBN and investigated how this affects the intercalation process. In Figure S10b and e, two CrSBr bulk flakes are shown. The flake in Figure S10b is fully encapsulated in hBN, while Figure S10e is only partially encapsulated with one end of the flake sticking out along the $a$-direction. Both heterostructures were fabricated on the same chip and intercalated together at the same time. After 38 min of intercalation the first flake is still in perfect condition without any change in appearance as shown in Figure S10c, and no change in the Raman signal shown in Figure S10a before and after the treatment (the difference in relative peak intensity is due to a difference in laser polarization angel during the measurement). From this, we can conclude that full hBN encapsulation is able to protect CrSBr from intercalation. No Li was able to penetrate through the hBN in the put of plane direction or slip underneath to reach the CrSBr flake.

The second flake on the other hand is strongly intercalated especially on the edges but throughout the whole length of the sheet in the $a$-direction as seen in Figure S10f and confirmed by the before and after Raman measurement in Figure S10d. From this, we can conclude that if Li is able to enter a flake it will easily diffuse within the flake along the $a$-direction and spread even underneath the hBN cover. This already serves as an indication that $a$ is the fast direction of diffusion.

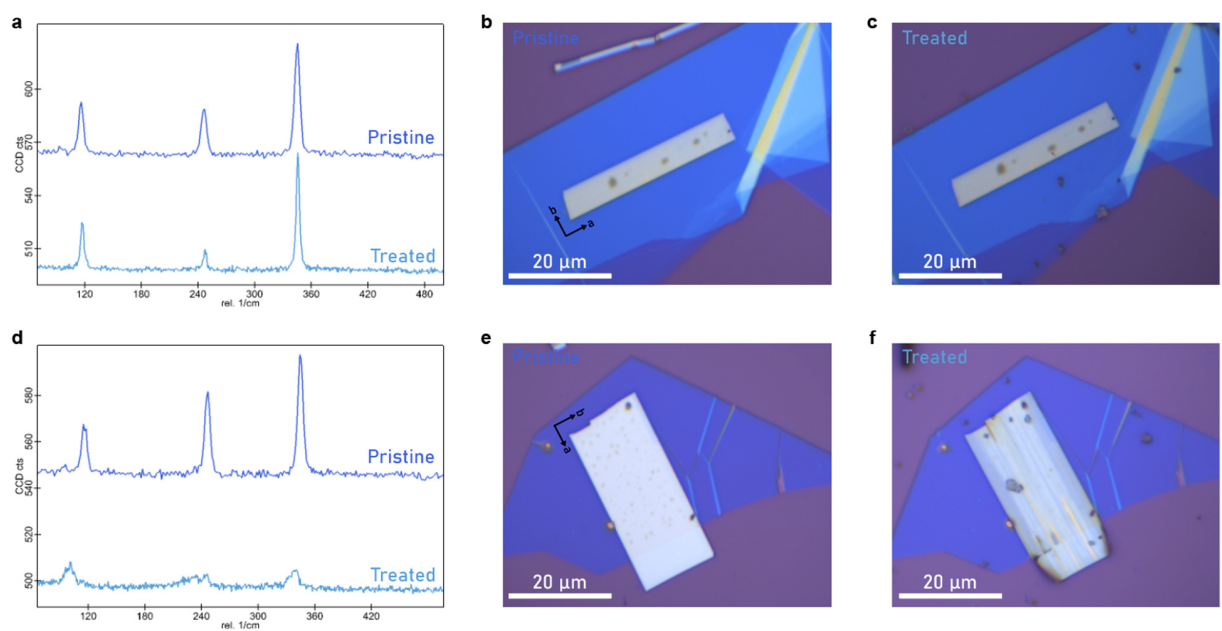

**Figure S10**: (a) Raman measurement before and after intercalation. (b) Pristine CrSBr flake, fully encapsulated in hBN. (c) Flake after treatment, with no visible signs of intercalation. (d) Raman measurement before and after intercalation (taken at the edge of the flake). (e) Pristine CrSBr flake, partially encapsulated in hBN (along *a*-direction). (f) Flake after treatment, with visible signs of intercalation along the whole length of the flake.

In Figure S11a, another CrSBr flake is shown, again partially covered by hBN, however, this time the part of it was uncovered along the *b*-direction. During 2h of intercalation we observed the intercalation progress slowly into the covered part of the sheet originating from the uncovered area, however in contrast to the previous experiment we do not see the Li diffuse quickly deep into the sheet. Instead, we observed a slow and gradual process with a sharp border between the intercalated and non-intercalated area. In Figure S11b, the flake is shown before and after the intercalation together with a Raman map indicating that the lower part of the flake was fully intercalated. The gradual progress during the intercalation is shown in Figure S11d.

The experiment was conducted with bottom contacted Au source and drain contacts to study how partially intercalating a flake will influence its electronic characteristics. Especially in the *b*-direction one might expect a semiconductor metal junction within the flake, which could lead to rectifying behavior or possibly even emission of photons. However, in our experiment we did not observe any noteworthy phenomena except for a slight decrease in conductivity likely due to the formation of slight cracks in the intercalated area. Nevertheless, the fact that we achieve partial intercalation in such a controlled manner opens up various possibilities for future experiments not only on fully, but also precisely partially intercalated devices.

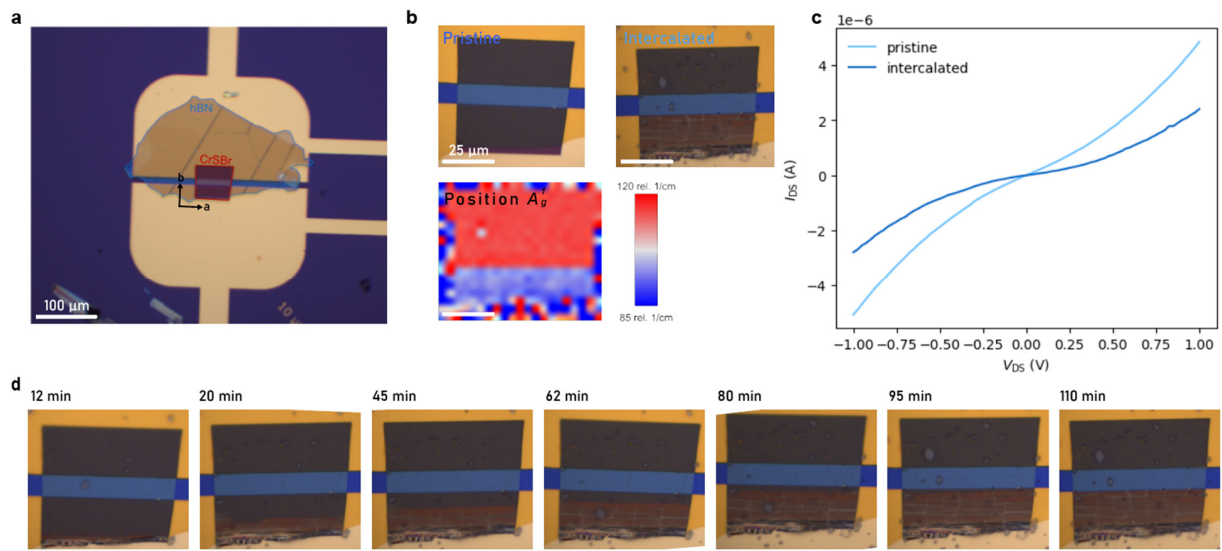

**Figure S11**: (a) Pristine flake of CrSBr partially encapsulated in hBN (along *b*-direction) bottom contacted with gold contacts. (b) Comparison of pristine and partially intercalated flake after 2h of intercalation. Intercalation is visible up to the middle of the flake as confirmed by Raman map. (c) IV sweep of pristine and intercalated flake. Except for a slight decrease in current, no major change was observed. (d) Gradual progression of intercalation along the *b*-direction over time.

## Supplementary Section 5. Diffusion Experiments.

The intercalation length ($L$) at different times $t$ was measured in ImageJ and plotted in Figure S12 against $t$. Two measurements in either direction per timestep were used to get a more accurate result. A crease in the top hBN layer was used as a reference point for the diffusion length for the diffusion in the *a*-direction. This assumes that the crease acts as a bottleneck for the diffusion into the rest of the sheet, with a constant concentration of Li$^+$ before the crease, allowing us to study the direction dependent diffusion kinetics beyond the crease. For the diffusion in the *b*-direction, which branches off perpendicularly from the diffusion path in the *a*-direction, the diffusion length from the reference point was considered the starting value for the diffusion in the *b*-direction.

Perfect 1D diffusion can be described by the equation $L^2 = 2Dt$ (here $L^2$ is equivalent to the mean squared displacement (MSD)). Thus, we expect a linear relation between $L^2$ and $t$. However, the data shows two distinct regimes for all the datasets: a slow diffusion at first, followed by a faster one. We fit both regimes of $L^2$ vs. $t$ for each dataset with a linear function and obtain two diffusion coefficients for each dataset from the slope of each fit $D = \frac{dL^2}{2dt}$. All fits are shown in Figure S9b-e with 95% confidence intervals for the fit. The measurement error was set as 0.513 µm, which was the approximate uncertainty when measuring the diffusion length from the optical micrograph. Overall, the fit works reasonably well, the sudden change in diffusion coefficient mid experiment is likely due to a sudden change in cell parameters, which allows Li$^+$ to diffuse faster. Such sudden stepwise changes in diffusion are also observed in other materials.[8]

While this is likely not a perfect description of the process, because of how $L_0$ was set, we can still use it to get an approximate value for the diffusion coefficients in each direction and calculate an anisotropy ratio which can later be compared with theoretical predictions. From the fits we calculate and average diffusion coefficient in either direction of $D_{a,1} = 9.0 \cdot 10^{-9}$ cm$^2$s$^{-1}$ and $D_{b,1} = 4.5 \cdot 10^{-10}$ cm$^2$s$^{-1}$ for the first regime and $D_{a,2} = 2.9 \cdot 10^{-8}$ cm$^2$s$^{-1}$ and $D_{b,2} = 1.2 \cdot 10^{-9}$ cm$^2$s$^{-1}$ for the second regime. By calculating the anisotropy ratio for each regime, we calculate an average of $\frac{D_a}{D_b} \approx 21.8 \pm 1.1$. The value for the $b$-direction is close to the reported Li ion diffusion coefficient in graphene of $D = 1.12 \cdot 10^{-10}$ cm$^2$s$^{-1}$ at 25°C and 0 state of charge.[9] The ratio of anisotropy is close to other anisotropic materials such as GeP, where the ratio is 25.[8]

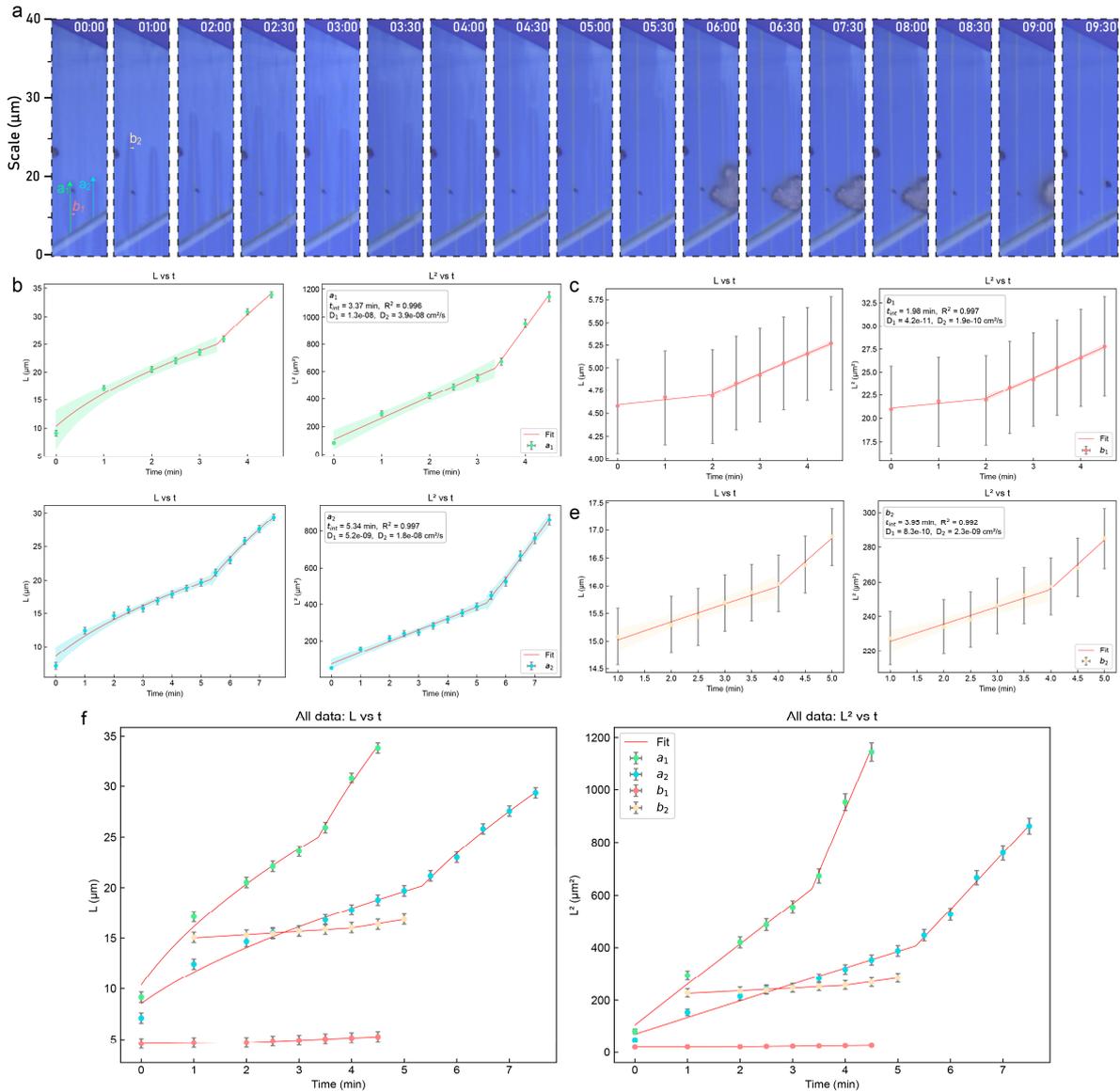

**Figure S12.** (a) Intercalation process within covered flake without annotation. (b-e) Fit for each dataset. Left: $L$ vs. $t$, right: $L^2$ vs. $t$. For regimes with 4 data points or more the 95% confidence interval is shown. Each datapoint is shown with error bars corresponding to a measurement uncertainty of 0.3 micron and 1s. (f) Combined data without subtraction of $L_0$ and $t_0$.

## Supplementary Section 6. Li diffusion from *ab initio* simulations.

1. Li intercalation

To simulate Li-intercalated CrSBr, we use a (3 × 2 × 2) supercell based on the entry 175897 in the ICSD database[10] (see also entry mp-22998 in the Materials Project database[11]), containing 72 atoms per cell with the composition $Cr_{24}S_{24}Br_{24}$. Li atoms are placed at the remaining *2b* (Cr) sites. We introduce 8 Li atoms into the supercell, yielding the composition $Li_8Cr_{24}S_{24}Br_{24}$, which corresponds to 11% Li intercalation in CrSBr. Figure S13a-c shows the pristine CrSBr (3 × 2 × 2) supercell, while Figure S13d-f depicts the Li-intercalated supercell, where 8 of the 24 available *2b* sites are occupied by Li.

Atomic positions and lattice vectors are relaxed using the Broyden–Fletcher–Goldfarb–Shanno (BFGS) algorithm as implemented in the QUANTUM ESPRESSO package.[12] We adopt the PBEsol exchange-correlation functional[13], include van der Waals interactions via Grimme's DFT-D3 dispersion correction[14], and perform spin-polarized calculations due to the presence of Cr atoms. After testing various k-point samplings[15], we select Γ-point-only sampling, given the large size of the supercell.

Table S1 reports the lattice parameters and cell angles of the relaxed Li-intercalated CrSBr (3 × 2 × 2) supercell. For comparison, we also include the relaxed lattice parameters and angles for a (1 × 1 × 1) CrSBr cell, calculated using a (3 × 2 × 2) Monkhorst–Pack k-point grid[15], alongside corresponding experimental values.[16] Li intercalation leads to structural changes at the Li sites, including a reduction in *z*-buckling within the Cr–S–Cr–S chains along the *y*-direction (Figure S13), accompanied by a 4% decrease in the *c*-axis and a 3% increase in the *b*-axis (Table S1).

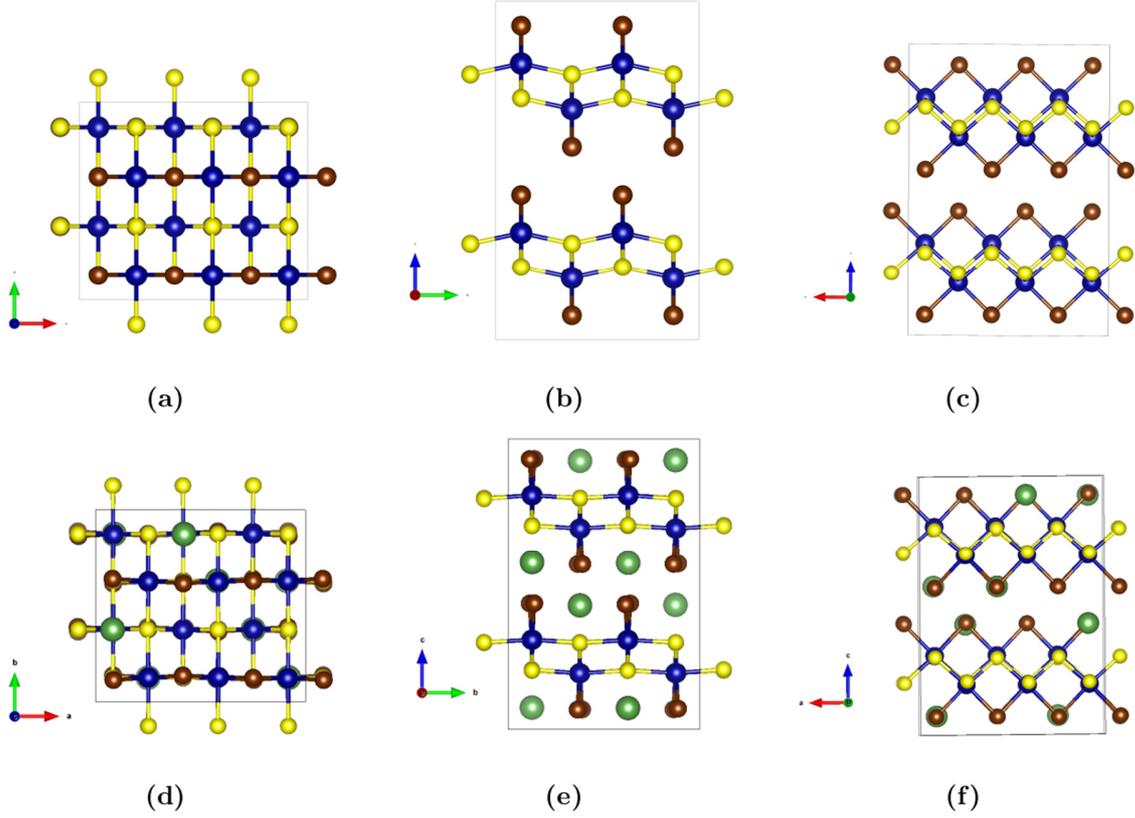

**Figure S13.** Pristine (a, b, c) and Li-intercalated (d, e, f) (3 × 2 × 2) supercells of CrSBr, containing 72 atoms (formula: $Cr_{24}S_{24}Br_{24}$) and 80 atoms (formula: $Li_8Cr_{24}S_{24}Br_{24}$), respectively; (a) and (d): top views (*ab* plane); (b) and (e): side views (*bc* plane), (c) and (f): side views (*ac* plane). Li atoms are displayed in green, Cr atoms in blue, S atoms in yellow, and Br atoms in brown. Images are generated by VESTA software.[17]

|  | CrSBr | CrSBr (experiments [7]) | LiCrSBr (3 × 2 × 2) |
|---|---|---|---|
| a (Å) | 3.56 | 3.50 | 10.53 |
| b (Å) | 4.71 | 4.74 | 9.74 |
| c (Å) | 7.75 | 7.91 | 14.88 |
| $\alpha$ (deg) | 90.0 | 90.0 | 89.96 |
| $\beta$ (deg) | 90.0 | 90.0 | 90.08 |
| $\gamma$ (deg) | 90.0 | 90.0 | 90.02 |

**Table S1.** Cell geometry (lattice parameters and angles) of the fully relaxed (1 × 1 × 1) cell of CrSBr (6 atoms per cell, formula: $Cr_2S_2Br_2$) using a (3 × 2 × 2) k-points Monkhorst-Pack grid[15], together with the corresponding experimental values[16] (first two columns, respectively). In the third column, the lattice parameters and angles of the LiCrSBr (3 × 2 × 2) supercell with 8 Li atoms

(formula: $Li_8Cr_{24}S_{24}Br_{24}$, see Fig. S13), fully relaxed at Γ, are reported. The latter is the supercell used for the AIMD simulations reported in this work.

2. Details of the *ab initio* Molecular Dynamics (AIMD) simulations

We perform *ab initio* Born–Oppenheimer molecular dynamics (AIMD) simulations[18], based on Kohn–Sham DFT[19,20] within the plane-wave pseudopotential framework[18,21], as implemented in the PW code of the QUANTUM ESPRESSO package.[12] Simulations are carried out in the canonical (*NVT*) ensemble using a stochastic velocity rescaling thermostat.[22]

We focus on four elevated temperatures (600, 800, 1000, and 1200 K), aiming to extrapolate Li diffusion down to room temperature. To account for the finite-temperature effects of Li intercalation on the cell volume and shape, we first perform a variable-cell (*NPT*) simulation at each temperature for ~10 ps. The thermally equilibrated supercells obtained from these *NPT* runs are then used in the subsequent *NVT* simulations. Figure S14 shows the total energies and instantaneous temperatures of the $Li_8Cr_{24}S_{24}Br_{24}$ supercell during the *NVT* dynamics.

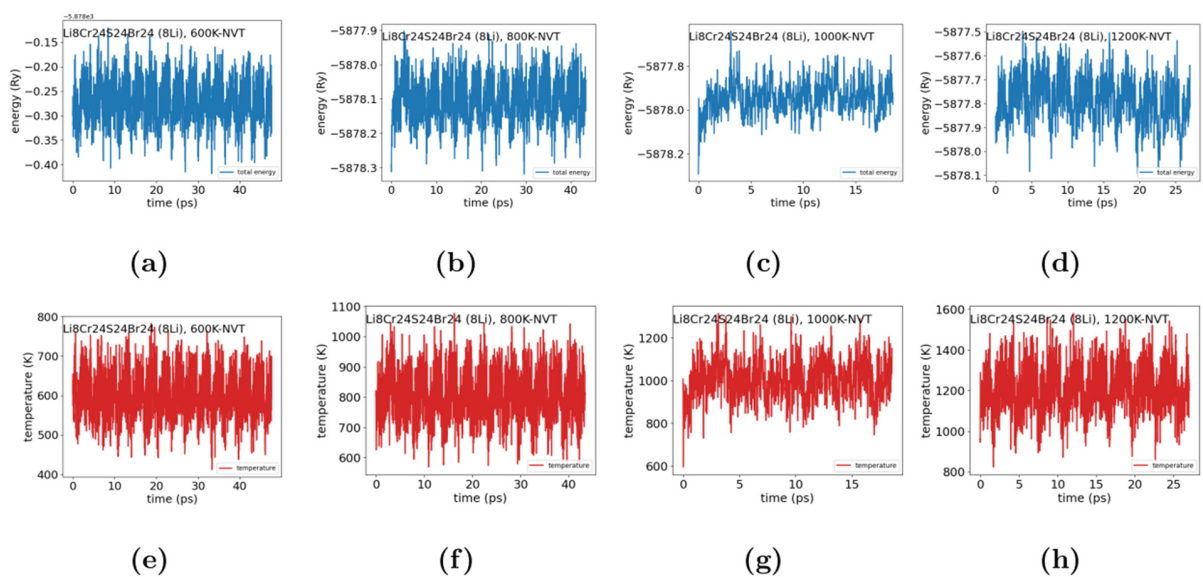

**Figure S14.** Energies (a-d) and temperatures (e-h) for $Li_8Cr_{24}S_{24}Br_{24}$ in the *NVT* simulations: a), e): 600 K; b), f): 800 K; c), g): 1000 K; d), h): 1200 K.

3. Li trajectories

In Figure S15a-b, we show the initial (a) and final (b) snapshots of the simulation at 600 K, projected onto the *ab* (top), *bc* (center), and *ac* (bottom) planes. The trajectory lines highlight the paths of Li ions and illustrate Li mobility, revealing a pronounced diffusion anisotropy along the *a*-direction. We verified that no other species exhibit mobility over the temperature range studied (see Figure S16). Supplementary Videos S1a-b, S2a-b, and S3a-b show the AIMD trajectories at 600, 800, and 1000 K, respectively (a: projections on the *xz*-plane; b: projections on the *yz*-plane).

Li atoms are depicted in green, each accompanied by a black trajectory line representing its path throughout the simulation. These visualizations clearly capture the diffusive behavior of the Li atoms, as well as the occurrence of individual atomic jumps during the dynamics. Li atoms easily jump on straight lines (channels) along the *x*-direction, with rarer off-channel jumps following a zig-zag pattern along *yz* that become more important at higher temperatures (800 and 1000 K).

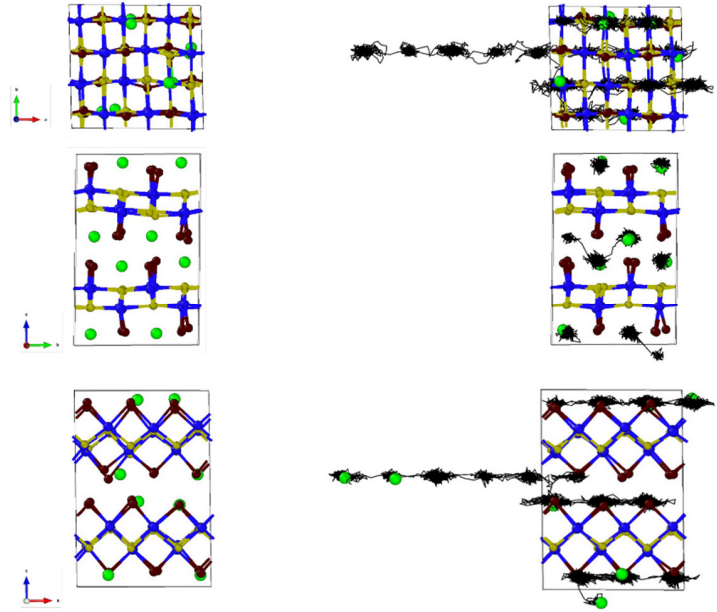

**Figure S15.** (a) The 11% Li-intercalated (3 x 2 x 2) supercell of CrSBr ($Li_8Cr_{24}S_{24}Br_{24}$) used for the AIMD simulations, projected on the *ab* (top), *bc* (center), and *ac* (bottom) planes. Li atoms are displayed in green, Cr atoms in blue, S atoms in yellow, and Br atoms in brown. (b) Final snapshot (t ~ 40 ps) of the $Li_8Cr_{24}S_{24}Br_{24}$ supercell, projected on the *ab* (top), *bc* (center), and *ac* (bottom) planes, from the 600 K AIMD simulations. Trajectory lines are displayed to visualize the Li paths. Images are obtained by using OVITO.[15]

4. Li mean squared displacements and diffusion coefficients

The mean squared displacement (MSD) of a given species provides a quantitative measure of its mobility within the material and is directly related to the self-diffusion coefficient. According to the Einstein relation for diffusion[23–25], the self-diffusion coefficient for the Li species in three dimensions is given by:

$$Di^{Li} = \lim_{t \to \infty} \frac{1}{6} \frac{d}{dt} \mathrm{MSD}^{Li}(t) \qquad (S3)$$

where the Li mean squared displacement $\mathrm{MSD}^{Li}(t)$ for a system of $N^{Li}$ atoms beyond the ballistic regime and over a sufficiently long time *t* is[25]:

$$\text{MSD}^{\text{Li}}(t) = \frac{1}{N_{\text{Li}}} \sum_i^{N_{\text{Li}}} |\mathbf{R}_i(t) - \mathbf{R}_i(0)|^2 \tag{S4}$$

Here, $\mathbf{R}_i(t)$ is the instantaneous position of the $i$-th Li atom, and $|\mathbf{R}_i(t) - \mathbf{R}_i(0)|^2$ is the squared displacement from its initial position.

Figure S16a–d presents $\text{MSD}^{\text{Li}}(t)$ as computed from Eq. S4, alongside the MSDs for the other species—Cr, S, and Br—denoted as $\text{MSD}^{\text{Cr}}(t)$, $\text{MSD}^{\text{S}}(t)$, $\text{MSD}^{\text{Br}}(t)$, respectively. These results confirm that Li is the only mobile species across all temperatures considered.

Figure S16e–h shows both $\text{MSD}^{\text{Li}}(t)$ and the individual squared displacements $|\mathbf{R}_i(t) - \mathbf{R}_i(0)|^2$ for each Li atom in the simulation cell, at 600, 800, 1000, and 1200 K.

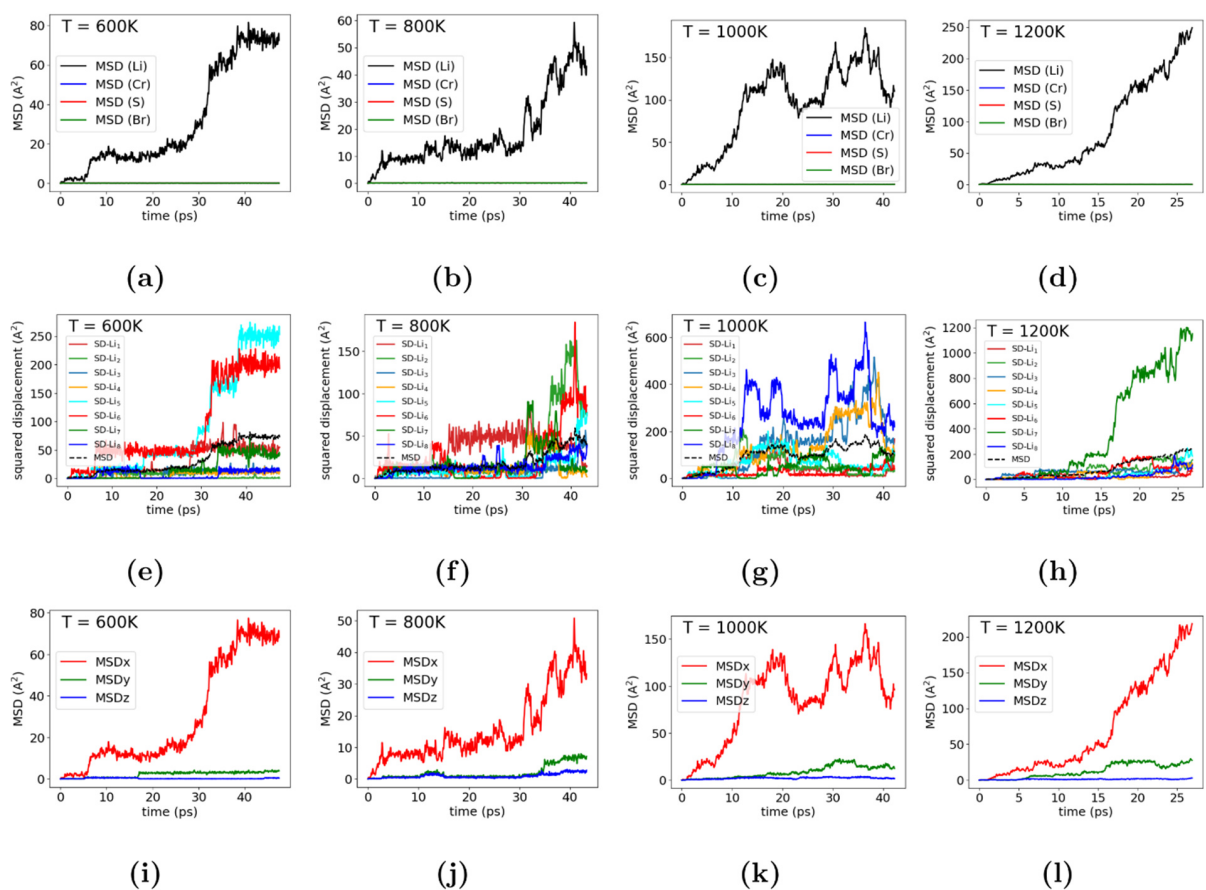

**Figure S16.** (a), (b), (c), (d): Mean squared displacements of the Li, Cr, S, and Br species in $\text{Li}_8\text{Cr}_{24}\text{S}_{24}\text{Br}_{24}$ derived from Eq. S4 at 600, 800, 1000, and 1200 K, respectively; (e), (f), (g), (h): Squared displacements of the single Li atoms in $\text{Li}_8\text{Cr}_{24}\text{S}_{24}\text{Br}_{24}$ at 600, 800, 1000, and 1200 K, respectively; (i), (j), (k), (l): x, y, and z components of the Li mean square displacement in $\text{Li}_8\text{Cr}_{24}\text{S}_{24}\text{Br}_{24}$ at 600, 800, 1000, and 1200 K, respectively.

Although AIMD simulations are highly accurate, they are limited to short time scales and small supercells[18], thus, even in diffusive regimes, they cannot provide a reliable diffusion coefficient from Eq.S3. Consequently, it is common practice to replace Eq.S3 with the time-averaged formulation of the Einstein relation[25–27]:

$$Di^{Li} = \lim_{t \to \infty} \frac{1}{6} \frac{d}{dt} \langle \text{MSD}^{Li}(t) \rangle \tag{S5}$$

Here, $\langle \text{MSD}^{Li}(t) \rangle$ is obtained by performing a statistical average over initial times $t'$ in the trajectory, leading to the following expression for the mean squared displacement:

$$\langle \text{MSD}^{Li}(t) \rangle = \frac{1}{N_{Li}} \sum_{i}^{N_{Li}} \langle |\mathbf{R}_i(t'+t) - \mathbf{R}_i(t')|^2 \rangle \tag{S6}$$

We report the averaged $\langle \text{MSD}^{Li}(t) \rangle$ from the simulations at 600, 800, 1000, and 1200 K in Figure S17. To evaluate the anisotropy of Li diffusion quantitatively, we extract the $x$-, $y$-, and $z$-components of $\langle \text{MSD}^{Li}(t) \rangle$ from the AIMD trajectories, which are shown in Figure S18a–c, d–f, g–i, and j–l for 600, 800, 1000, and 1200 K, respectively. Figure S18a–b shows a $D_x^{Li}/D_y^{Li}$ ratio (with $D_x^{Li}$ and $D_y^{Li}$ derived from Eq. S5 of 20.8 ± 7.2, in good agreement with the experimental results, as well as a $D_x^{Li}/D_z^{Li}$ ratio of approximately 40. A pronounced temperature dependence of the $D_x^{Li}/D_y^{Li}$ ratio is observed across the datasets: the ratio decreases from 20.8 ± 7.2 at 600 K to 6.5 ± 3.3 at 800 K, 5.9 ± 2.2 at 1000 K, and 4.4 ± 1.2 at 1200 K (Figure S18). Interestingly, the most significant reduction occurs between 600 and 800 K, suggesting the existence of two distinct diffusion regimes: one below 600–800 K, where Li predominantly migrates through $x$-oriented channels, and one above 600–800 K, where diffusion through $y$-oriented channels becomes accessible.

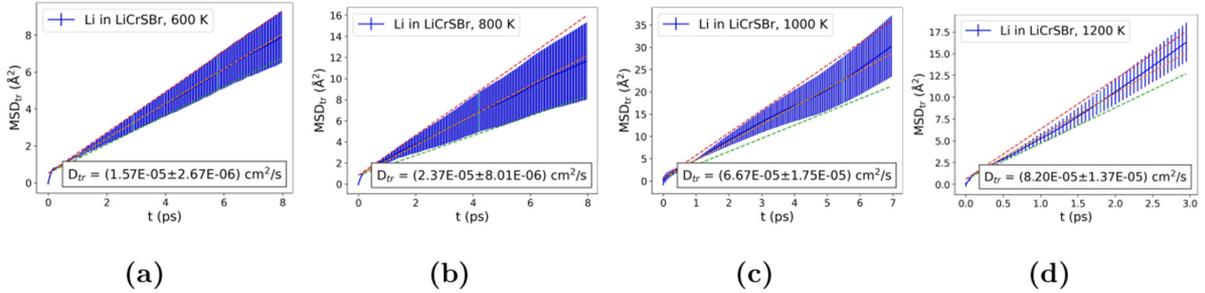

(a)          (b)          (c)          (d)

**Figure S17.** Li mean squared displacements averaged over the initial times (Eq. S6), from the 600 K (a), 800 K (b), 1000 K (c), and 1200 K (d) simulations. In each plot, the diffusion coefficient at that temperature derived from Eq. S5 is reported in the bottom panel.

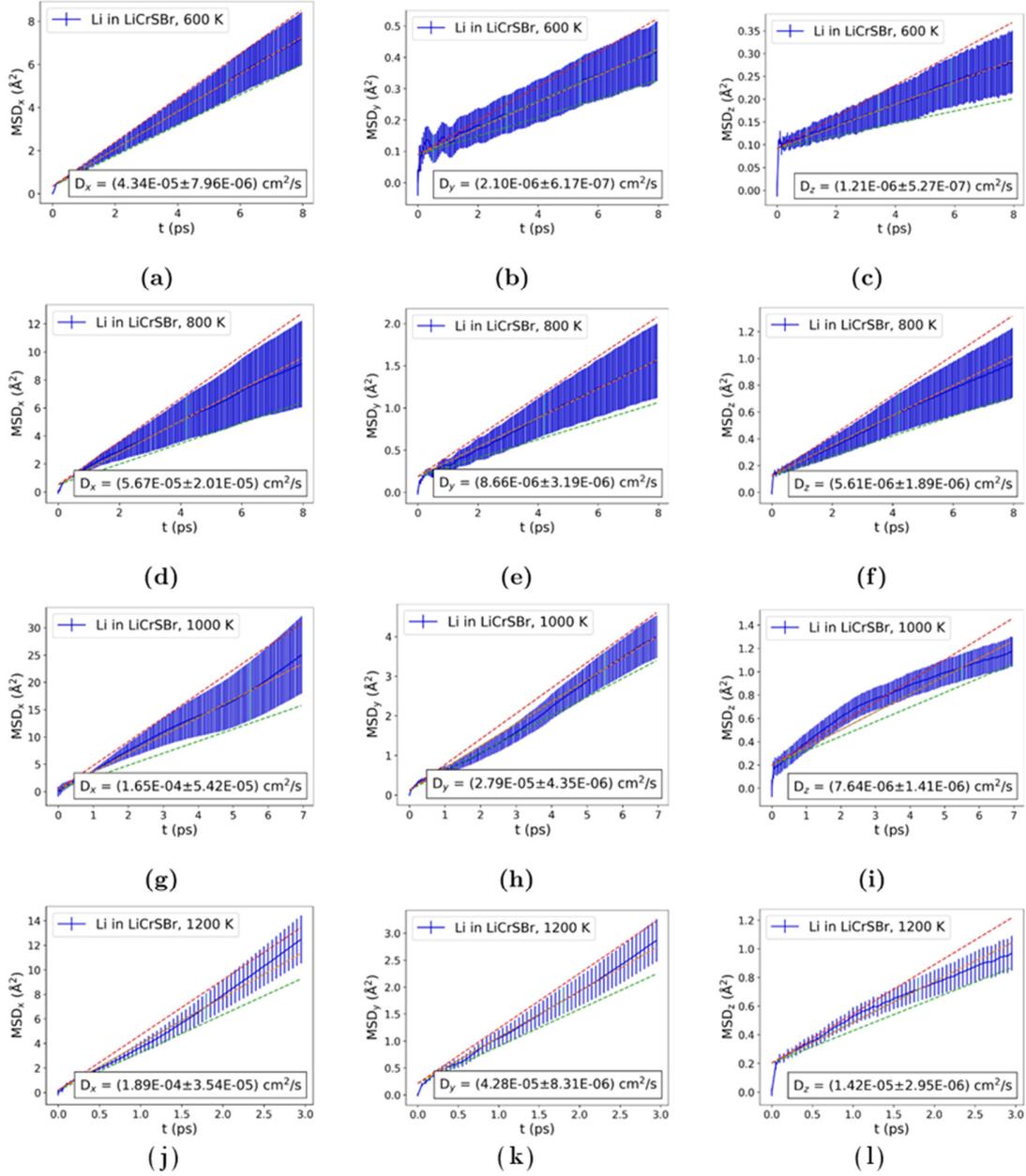

**Figure S18.** Components along the $x$, $y$, and $z$-directions of the Li mean square displacement averaged over the initial times (Eq. S4), from the AIMD simulations at 600 K (a, b, and c, respectively), 800 K (d, e, and f, respectively), 1000 K (g, h, and i, respectively), and 1200 K (j, k, and l, respectively). In each plot, the direction-resolved diffusion coefficient at that temperature is reported in the bottom panel. From these, we extract values for the $D_x/D_y$ ratio of 20.8, 6.5, 5.9, and 4.4 at 600, 800, 1000, and 1200 K, respectively.

Finally, we treat Li diffusion as an activated process that follows an Arrhenius-type behavior[28]:

$$\ln D^{\text{Li}}(T) = \ln A - \frac{E_{a_{D^{\text{Li}}}}}{k_B T} \tag{S7}$$

In this expression, $A$ is a pre-exponential factor related to the attempt frequency, and $E_{aD}{}^{Li}$ represents the activation energy barrier for Li diffusion. The Arrhenius plot for $D^{Li}$ is presented in Figure S19 is in the conventional form $\log_{10} D^{Li}$ versus $1000/T$. From this analysis, we estimate the extrapolated Li diffusion coefficient at 300 K to be approximately $6.0 \times 10^{-7}$ cm$^2$s$^{-1}$.

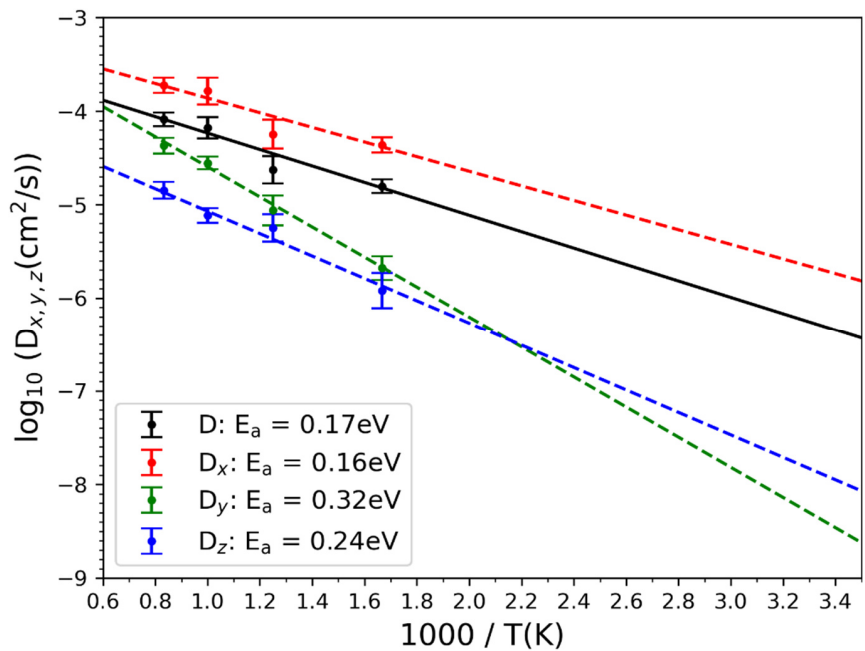

**Figure S19.** Arrhenius plot for Li diffusivity in Li$_8$Cr$_{24}$S$_{24}$Br$_{24}$ from the AIMD simulations, with the $x$, $y$, and $z$ components of the diffusivity. Below 600 K, lines are drawn to obtain an estimate of the diffusion coefficient at 300 K (~ $6.0 \times 10^{-7}$cm$^2$s$^{-1}$). Corresponding activation energies are presented in the inset.

## Supplementary Section 7. Additional Time Dependent Conductivity Measurement.

Figure S20 shows a characterization of a simple two-terminal device based on a few layered sheets of CrSBr (~9 layers, see Figure 20b). Figure 20a shows the device before and after intercalation as well as after 15 h of exposure to ambient conditions. During the intercalation, a slight change in color becomes visible, and after 15 h on air, small dark spots can be seen at the device edges, which are likely caused by lithium exiting the sheet and forming lithium salts when reacting with air. The Raman measurements shown in Figure 20c confirm the intercalation of the sheet indicated by the emergence of $A_1^1$, $A_1^2$, and $A_1^3$ modes at around 17 cm$^{-1}$, 16 cm$^{-1}$, and 7 cm$^{-1}$ lower energies than the pristine modes respectively. During 15 h in ambient conditions, the original peaks return, indicating a slow deintercalation, however, they never quite reach the position of the non-intercalated material. An offset of 1.5 cm$^{-1}$ remains, as well as a slight shoulder of the $A_g^2$ peak, at this point the sheet is most likely still slightly intercalated but to a much lesser extent as it had been initially. In this device substantial contact resistance and Schottky behavior was observed even in the pristine measurement, as a result the increase in conductivity is higher than in the device in Figure 5 of the main text, around one order of magnitude. In fact, the contacts appear to be improving, indicated by the higher linearity and steeper curve in the low bias regime. In the high bias regime, we still observe the same trend of gradual decrease in conductivity associated with the deintercalation of Li and corresponding reduction in charge carrier density.

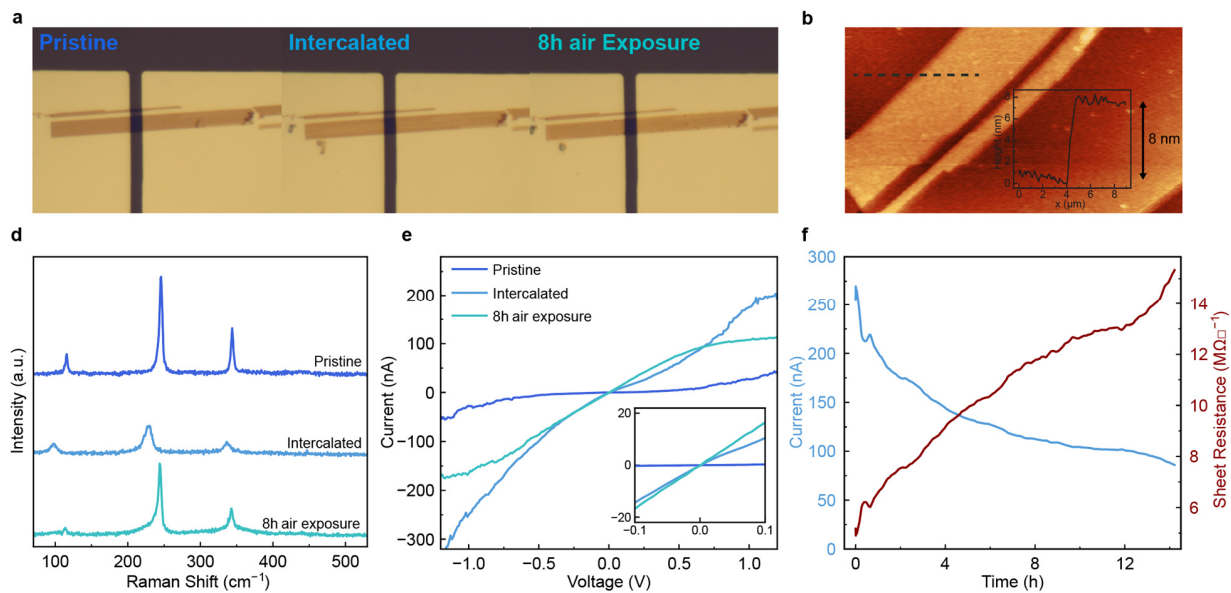

**Figure S20**: (a) Optical micrograph of one CrSBr device in a pristine state, intercalated, and after exposure to ambient conditions for 15 h. Scale bar 15 μm. (b) AFM plot of the CrSBr device with height profile as an inset. The thickness of the devices is ca. 8 nm corresponding to 9 layers. The scale bar is 5 μm. (c) Raman spectra of the device in a pristine state, intercalated, and after exposure to ambient conditions for 15 h. (d) Output characteristics of the device in a pristine state, intercalated, and after exposure to ambient conditions for 15 h. The inset shows the highly linear low bias regime. (e) Continuous conductivity measurement during the exposure to ambient conditions with $V_{DS} = 1$ V. The left axis shows the current during the measurement and the right the sheet resistance $R_s$.

## Supplementary Section 8. Temperature Dependence of Conductivity.

Figure S21a shows the temperature dependence of conductivity of a few-layer sheet of Li-CrSBr in the *a* and *b*-direction and for reference also the *a*-direction of pristine CrSBr. The raised Fermi-level due to the extra charge carriers in Li-CrSBr should theoretically lead to metallic behavior at least in the *b*-direction[29]. We observe an increase in conductivity by more than one order of magnitude for both directions; however, we still see thermally activated transport. Notably, we also observe a substantially faster drop in conductivity with decreasing temperature, compared to the pristine case, in all our measurements. It had previously been reported that bulk Li-CrSBr shows a feature around 200 K which suggests the onset of ferromagnetic ordering. As our initial measurement shows a measurement artefact (related to switching of the preamplifier) in this region, we repeated the measurement with a two terminal device only along the *a*-direction. As shown in Figure S21b this device shows a similarly steep decrease in conductivity, compared to the pristine device, but as shown in the inset there is a noticeable feature at around 235 K, which could be related to ferromagnetic ordering. Since we observe the feature both during the heating and cooling, it is likely a real feature, although it is 30 K higher than expected.

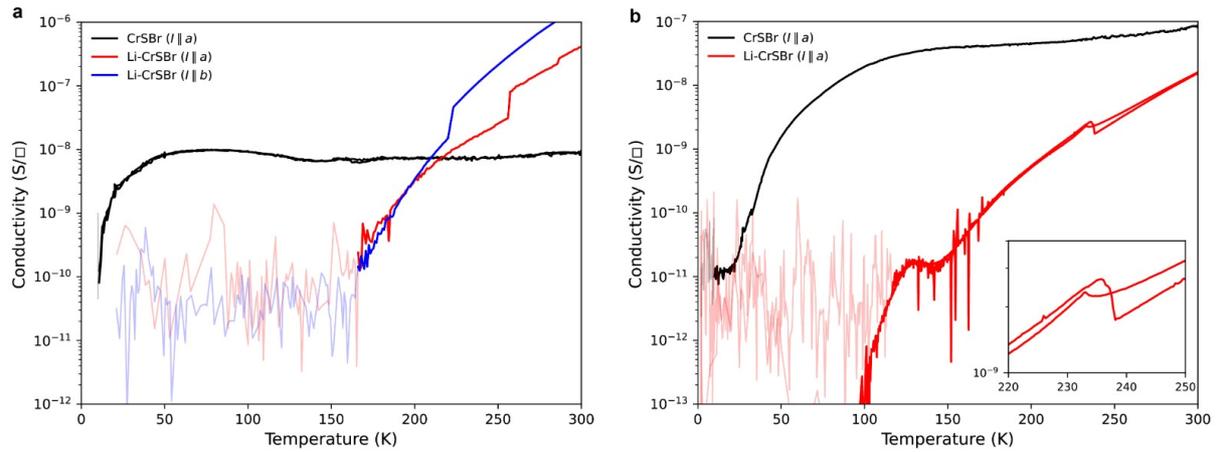

**Figure S21**. (a) Temperature dependence of CrSBr (*a*-direction) compared to Li-CrSBr (*a* and *b*-direction). Measured on the same sample (two sets of perpendicular contacts were used to contact the sheet). Room temperature conductivity is higher than in the pristine flake but drops fast with decreasing temperature. The sudden change between 200 and 250 K is a measurement artifact from switching the preamplifier during the measurement. Semi-transparent curves correspond to the noise floor. (b) Temperature dependence of CrSBr (*a*-direction) compared to Li-CrSBr (*a*-direction) measured on a different device. A simpler two contact geometry was used. A similar decrease in conductivity is observed. A feature is visible around 235 K which is likely related to ferromagnetic ordering. The inset shows the feature region more closely. Semi-transparent curves correspond to the noise floor.